\documentclass[aps,prxquantum,reprint,superscriptaddress]{revtex4-2}

\usepackage{amsmath,amssymb,amsthm}
\usepackage{graphicx}
\usepackage{dcolumn}
\usepackage{physics}
\usepackage{lipsum}
\usepackage{cleveref}
\usepackage[super]{nth}
\usepackage{xcolor}
\usepackage{quantikz}

\usepackage[compat=0.6]{yquant}
\useyquantlanguage{groups}
\usepackage{tikz}
\usetikzlibrary{positioning}
\usetikzlibrary{graphs}
\yquantset{
   operators/every box/.append style={/yquant/operators/every rectangular box}, 
   operators/every control box/.style={/yquant/operators/every rectangular box}
}
\makeatletter
\yquant@langhelper@declare@command@uncontrolled%
   {controlbox}%
   \yquant@register@get@allowmultitrue%
   {%
      \expandafter\yquant@prepare@injection%
         \expandafter{\yquant@lang@attr@value}%
         {/yquant/operators/every control box}%
   }
\makeatother

\usepackage{algorithm}
\usepackage{algpseudocode}

\makeatletter
\newenvironment{breakablealgorithm}
  {
   \begin{center}
     \refstepcounter{algorithm}
     \hrule height.8pt depth0pt \kern2pt
     \renewcommand{\caption}[2][\relax]{
       {\raggedright\textbf{\ALG@name~\thealgorithm} ##2\par}%
       \ifx\relax##1\relax 
         \addcontentsline{loa}{algorithm}{\protect\numberline{\thealgorithm}##2}%
       \else 
         \addcontentsline{loa}{algorithm}{\protect\numberline{\thealgorithm}##1}%
       \fi
       \kern2pt\hrule\kern2pt
     }
  }{
     \kern2pt\hrule\relax
   \end{center}
  }
\makeatother

\usepackage{subcaption}

\usepackage{pgfplots}
\usepackage{multirow}
\usepackage{xparse}
\usepackage[inline,shortlabels]{enumitem}
\usetikzlibrary{angles,3d,intersections}
\usepackage{capt-of}
\newtheorem{theorem}{Theorem}
\newtheorem{definition}{Definition}

\newtheorem{lemma}{Lemma}
\newtheorem{corollary}{Corollary}

\newcounter{mythm}       
\newcounter{mydef}       
\newcounter{mylem}       
\newcounter{myconj}      
\newcounter{mycoro}

\renewenvironment{theorem}[1][]
  {\refstepcounter{mythm}\par
   \textit{{Theorem~\themythm}#1---}\nopagebreak\noindent\ignorespaces}
  {\par}

\renewenvironment{definition}[1][]
  {\refstepcounter{mydef}\par
   \textit{{Definition~\themydef}#1---}\nopagebreak\noindent\ignorespaces}
  {\par}

\renewenvironment{lemma}[1][]
  {\refstepcounter{mylem}\par
   \textit{{Lemma~\themylem}#1---}\nopagebreak\noindent\ignorespaces}
  {\par}

\renewenvironment{corollary}[1][]
  {\refstepcounter{mycoro}\par
   \textit{{Corollary~\themycoro}#1---}\nopagebreak\noindent\ignorespaces}
  {\par}

\newcommand{\hwp}{\ket{\psi_{\text{H}}}}

\newcommand{\C}{\mathbb{C}}
\newcommand{\Span}{\operatorname{span}}
\newcommand{\End}{\operatorname{End}}

\allowdisplaybreaks

\begin{document}

\preprint{preprint]}

\title{Preparation of Hamming-Weight-Preserving Quantum States with Log-Depth Quantum Circuits}

\author{Yu Li}
\affiliation{%
State Key Lab of Processors,  Institute of Computing Technology,  Chinese Academy of Sciences, 
Beijing 100190,  China\\
}%
\affiliation{
School of Computer Science and Technology,  University of Chinese Academy of Sciences,  Beijing
100049,  China\\
}%

\author{Guojing Tian} \email{tianguojing@ict.ac.cn}
 \affiliation{%
State Key Lab of Processors,  Institute of Computing Technology,  Chinese Academy of Sciences, 
Beijing 100190,  China\\
}%
\affiliation{
School of Computer Science and Technology,  University of Chinese Academy of Sciences,  Beijing
100049,  China\\
}%

\author{Xiaoyu He}
\affiliation{%
State Key Lab of Processors,  Institute of Computing Technology,  Chinese Academy of Sciences, 
Beijing 100190,  China\\
}%
\affiliation{
School of Computer Science and Technology,  University of Chinese Academy of Sciences,  Beijing
100049,  China\\
}%

\author{Junhong Nie}
 \affiliation{%
Shandong University, Jinan 250100, China\\
}%

\author{Jiadong Zhu}
\affiliation{%
State Key Lab of Processors,  Institute of Computing Technology,  Chinese Academy of Sciences, 
Beijing 100190,  China\\
}%
\affiliation{
School of Computer Science and Technology,  University of Chinese Academy of Sciences,  Beijing
100049,  China\\
}%

\author{Xiaoming Sun} \email{sunxiaoming@ict.ac.cn}
\affiliation{%
State Key Lab of Processors,  Institute of Computing Technology,  Chinese Academy of Sciences, 
Beijing 100190,  China\\
}%
\affiliation{
School of Computer Science and Technology,  University of Chinese Academy of Sciences,  Beijing
100049,  China\\
}%

\date{\today}

\begin{abstract}

Hamming-Weight-Preserving (HWP) states, $\vert\psi_{\mathrm{H}}\rangle=\sum_{\mathrm{HW}(x)=k}\alpha_x\vert x\rangle$, represent a fundamental class of quantum states on $n$ qubits with fixed excitation number $k$, playing a pivotal role in quantum simulation, machine learning, and optimization, with prominent examples including the Bethe eigenstates of integrable models and Configuration Interaction states. However, existing preparation methodologies typically rely on sequential architectures whose depth scales linearly with the subspace dimension, whereas practical deployment on realistic quantum hardware demands shallow circuits to mitigate decoherence. In this work, we present quantum circuits that achieve the information theoretic limit of depth logarithmic in the number of populated computational basis states while maintaining asymptotically optimal size. Specifically, for general HWP states, we provide a construction that matches the lower bounds of $\Theta(\log \binom{n}{k})$ depth and $\Theta(\binom{n}{k})$ size. Furthermore, we propose a resource efficient scheme utilizing only $O(\log k)$ ancillary qubits. For the $k=2$ case, where HWP states naturally admit a graph representation, we show that states whose support graphs are trees or grids admit ancilla free preparation with $\Theta(\log n)$ depth and $\Theta(m)$ size. Overall, our work provides logarithmic depth state preparation algorithms across various regimes while establishing a connection between quantum state synthesis and Krawtchouk polynomials.
\end{abstract}
\maketitle


\section{Introduction}
Preparing structured quantum states is a foundational task for quantum
simulation, quantum machine learning, and quantum optimization. A particularly
versatile family is the Hamming-Weight-Preserving (HWP) states,
\begin{equation}\label{eq:hwp}
\ket{\psi_{\text{H}}} = \sum_{\text{HW}(x)=k} \alpha_x \ket{x},
\end{equation}
which span the fixed-excitation subspaces dictated by particle-number or
magnetization conservation. In quantum many-body physics, HWP states describe
systems with $U(1)$ symmetry~\cite{marvian2022restrictions} and contain the Bethe
eigenstates of integrable spin chains~\cite{bethe1931theorie,gaudin2014bethe}, making them
natural inputs for digital simulation of quantum magnetism, topological phases,
and many-body localization~\cite{vasseur2016nonequilibrium}. In quantum machine learning,
restricting variational ansatze to HWP sectors reduces the effective Hilbert
space and helps mitigate barren plateaus~\cite{monbroussou2025trainability}. In
combinatorial optimization, the HWP structure is well suited to
cardinality-constrained problems such as portfolio selection~\cite{he2023alignment}, and
in quantum error correction, constant-excitation codes built from HWP states
are intrinsically resilient to amplitude damping~\cite{chuang1997bosonic}. Important
special cases such as Dicke states have additional applications in quantum
metrology, game theory, and quantum
networking~\cite{dicke1954coherence,ozdemir2007necessary,prevedel2009experimental,yang2018quantum,su2020quantum}. Realizing these
applications on near-term and fault-tolerant hardwares therefore requires efficient preparation of HWP states, that is, prepare these states with circuits of lower depth and fewer gates.

A number of existing protocols can in principle be used to prepare HWP states.
Universal state preparation algorithms target arbitrary $n$-qubit states with
asymptotically optimal size, and additional ancillary qubits allow further
depth reduction~\cite{sun2023asymptotically,yuan2023optimal}. Sparse-state preparation is also
applicable, since an HWP state is supported on only $\binom{n}{k}$ basis
states, and the gate count then scales with the number of nonzero
amplitudes~\cite{gleinig2021efficient,mao2024towards,li2025nearly}. More recently,
constructions specialized to fixed-Hamming-Weight or Bethe-type states have
been developed: ancilla-free recursive schemes prepare HWP states with
$O(k\binom{n}{k})$ gates~\cite{raveh2024deterministic,farias2025quantum}, while subsequent work
reduces the size to nearly optimal or optimal scaling using ancillary
qubits~\cite{mao2024towards,luo2025optimal}. None of these methods, however, achieves
logarithmic depth and optimal circuit size at the same time. For special cases, Dicke states~\cite{dicke1954coherence}, cyclic states~\cite{mozafari2022efficient} and Antisymmetrized Geminal Power (AGP) states~\cite{ring2004nuclear,khamoshi2023state} have also been studied.
The complexity of existing algorithms for HWP states is summarized in Tab.~\ref{tab:comparison}. Universal
protocols ignore the fixed-Hamming-Weight structure, sparse-state algorithms
treat the allowed basis states as unstructured indices, and existing
HWP-specific constructions are essentially sequential in depth. Closing this
gap is especially relevant for near-term devices, where decoherence makes
circuit depth a critical resource.

\begin{table*}
    \centering
    \caption{Various quantum state preparation algorithms for HWP states and complexity results.}
    \label{tab:comparison}
    \renewcommand{\arraystretch}{1.5}
    
    \scalebox{0.7}{%
    \begin{tabular}{c|c|c|c|c|c}
\hline\hline
Preparation algorithm                & Excitations               & Reference                          & \# of ancillary qubits & \# of CNOT gates & Depth                                                                                  \\ \hline
\multirow{4}{*}{Sparse state preparation} & \multirow{4}{*}{Arbitrary $k$} & \cite{gleinig2021efficient}  & 0                          & $O(n {n \choose k})$ & ---                                                                 \\
~                            & ~                         & \cite{mao2024towards}        & 2                          & $O(\frac{n {n \choose k}}{\log n}+n)$ & ---                                                                      \\ 
~                            & ~                         & \cite{li2025nearly}          & $l \in O(\frac{n{n \choose k}}{\log n{n \choose k}}+n)$                           & $O\left(\frac{n{n \choose k}}{\log (l+n)}+n\right)$ & ---                                                                      \\ 
~                            & ~                         & \cite{luo2025space}          & $l \ge 6n$         & $O\left(\frac{n{n \choose k}}{\log (l/n)}+\frac{n{n \choose k}}{\log d}\right)$ & $O\left(\frac{n{n \choose k}\log l}{l \log (l/n)}+\log n{n \choose k}\right)$                                                      \\ 
\hline
\multirow{8}{*}{HWP state preparation} & $k=2$ (Graph-structured states with $m$ edges) & This work & $O(m)$ & $\Theta(m)$ & $\Theta(\log n)$  \\
~       & $k=2$ (Tree/Grid-structured states)                         & This work & 0 & $\Theta(n)$ & $O(\log n)$  \\ \cline{2-6}
~                   & \multirow{6}{*}{Arbitrary $k$} & \cite{raveh2024deterministic}\cite{farias2025quantum} & $0$ & $O\left(k{n \choose k}\right)$ & $O\left(k{n \choose k}\right)$\textsuperscript{*}  \\
~                   & ~                         & \cite{mao2024towards} & $O(n)$ & $O(\binom{n}{k} \log n)$ & $O(\binom{n}{k} \log n)$\textsuperscript{*}  \\ 
~                   & ~                         & \cite{luo2025optimal} & $O(n)$ & $\Theta(\binom{n}{k})$ & $O(\binom{n}{k})$\textsuperscript{*}  \\ 
~                   & ~                         & This work & $O\left({n \choose k}\right)$ & $\Theta\left({n \choose k}\right)$ & $\Theta\left(\log{{n \choose k}}\right)$  \\ 
~ & ~                  & This work & $O\left(\log{k}\right)$ & $\Theta\left({n \choose k}\right)$ & $O(\binom{n}{k})$ \\ \hline \hline
    \end{tabular}%
    }

    \vskip 2mm
    \begin{footnotesize}
        \begin{list}{}{\leftmargin=1.5em \labelwidth=1em \labelsep=0.5em}
          \item[*] These depth results were not explicitly provided in the original papers but were calculated by us following their described methods.
        \end{list}
    \end{footnotesize}
\end{table*}

In this work, we provide a complete complexity landscape for HWP state preparation by matching our protocols with rigorous lower bounds. For the general $k$ regime, we first provide a construction that matches the information-theoretic lower bounds of $\Theta(\log \binom{n}{k})$ depth and $\Theta(\binom{n}{k})$ size. To optimize spatial resources, we further propose a resource-efficient scheme requiring only $O(\log k)$ ancillary qubits while maintaining $\Theta(\binom{n}{k})$ gate complexity. Focusing on the $k=2$ case, which we map to undirected weighted graphs, we develop an algorithm achieving $\Theta(\log n)$ depth and $\Theta(m)$ size for any $m$-edge graph. Furthermore, for trees and grids, we demonstrate that the ancillary requirement can be entirely eliminated without sacrificing logarithmic depth. Notably, while this work focuses on the preparation of HWP states with real coefficients, the extension to complex amplitudes is straightforward. This generalization only requires adjusting the phase parameters of the two-qubit gates of each proposed algorithm. These results establish optimal primitives for HWP state preparation across both near-term and fault-tolerant quantum architectures.

The rest of this paper is organized as follows. In Sec. \ref{sec:pre}, we introduce the basic tools used throughout this work, including fan-in and fan-out gates, unary encoding, and the Hamming weight counter. To build intuition and ensure a smooth presentation, we will first introduce our constructions for the $k=2$ case, which serves as a foundation before generalizing the framework to arbitrary $k \ge 3$. In Sec. \ref{sec:k=2}, we study graph-structured states, covering the general case as well as tree- and grid-structured states. In Sec. \ref{sec:HWP}, we present our optimal-depth construction for general HWP state preparation. In Sec. \ref{sec:HWP fewer}, we develop an ancilla-efficient preparation scheme and establish the existence of the parameters required in its construction. In Sec. \ref{sec:lower}, we derive lower bounds on the circuit depth and size for HWP state preparation. In Sec. \ref{sec:numerical}, we present numerical results. Finally, Sec. \ref{sec:conclusion} concludes the paper and discusses several directions for future research.

\section{Preliminaries}\label{sec:pre}
Pauli gates and rotation gates about X-axis, Y-axis, and Z-axis are fundamental single-qubit gates in quantum computation. The CNOT gate is an elementary 2-qubit gate. The single-qubit gates, together with the CNOT gate, constitute a universal quantum gate set, thus they are usually referred to as the elementary quantum gates. In this article, when we talk about size complexity or depth complexity, we only consider circuits that consist of elementary quantum gates. We will frequently use the $X$ gate, the $H$ gate, the $C^2(R_y)$ gate and the $C^n(X)$ gate, where $C^2(X)$ is also known as the Toffoli gate. The matrix representations of $X$ gate,  $R_y(\theta)$ gate,  $H$ gate,  and CNOT gate are as follows.

\begin{equation}
\begin{aligned}
X=
\begin{pmatrix}
0 & 1 \\
1 & 0
\end{pmatrix}
, R_y(\theta)=
\begin{pmatrix}
\cos{\frac{\theta}{2}} & -\sin{\frac{\theta}{2}} \\
\sin{\frac{\theta}{2}} & \cos{\frac{\theta}{2}}
\end{pmatrix}, \\
H=\frac{1}{\sqrt{2}}
\begin{pmatrix}
1 & 1 \\
1 & -1
\end{pmatrix}
, \text{CNOT}=
\begin{pmatrix}
1 & 0 & 0 & 0 \\
0 & 1 & 0 & 0 \\
0 & 0 & 0 & 1 \\
0 & 0 & 1 & 0
\end{pmatrix}.
\end{aligned}
\end{equation}

Actually, $C^2(R_y)$ gates and $C^2(X)$ gates can be implemented by using single-qubit gates and CNOT gates as shown in Fig.~\ref{Decomp}.
\begin{figure}[h]
    \centering
    \begin{tikzpicture}
        \begin{yquant}[register/separation=2mm, operator/separation=0.1mm]
            qubit {} q0;
            qubit {} q1;
            qubit {} q2;

            box {$R_y(2\theta)$} q2 | q0, q1;

            text {\space =} (-);
            cnot q2 | q1;
            box {$R_y(-\frac{\theta}{2}$)} q2;
            cnot q2 | q0;
            box {$R_y(\frac{\theta}{2}$)} q2;
            cnot q2 | q1;
            box {$R_y(-\frac{\theta}{2}$)} q2;
            cnot q2 | q0;
            box {$R_y(\frac{\theta}{2}$)} q2;
        \end{yquant} 
    \end{tikzpicture}

    \vspace{0.5cm}
    
    \begin{tikzpicture}
        \begin{yquant}[register/separation=2mm]
            qubit {} q0';
            qubit {} q1';
            qubit {} q2';

            cnot q2' | q0', q1';

            text {\space  = \space } (-);
            box {$V$} q2' | q1';
            cnot q1' | q0';
            box {$V^\dagger$} q2' | q1';
            cnot q1' | q0';
            box {$V$} q2' | q0';
        \end{yquant}
    \end{tikzpicture}
    \caption{The decompositions of the $C^2 ( R_y(2\theta) )$ gate and Toffoli gate,  where $V=(1-i)(I+iX)/2$.}
    \label{Decomp}
\end{figure}

$C^n(X)$ gates can also be decomposed into elementary gates.

\begin{lemma} \label{lem:C^n(x)}
(\cite{Craig}) A $C^n(X)$ gate can be implemented by a quantum circuit of depth and size $O(n)$ with $1$ ancillary qubit.
\end{lemma}

Additionally, we give a simple way to calculate the depth and size complexity of any controlled gate, $C(U)$ when both the depth and size complexity of $U$ are known.

\begin{lemma} \label{lem:C(U)}
    Assume $U$ is an $n$-qubit quantum gate, the circuit implementing $U$ has $d$ depth and $s$ size. When using $O(s)$ ancillary qubits, $C(U)$ can be implemented with $O(\log s+d)$ depth and $O(s)$ size.
\end{lemma}

\noindent \textit{Proof.} By making $O(s)$ copies of the control qubit, all the elementary gates in the circuit implemented $U$ can get a unique control qubit, hence the elementary gates that are originally in the same layer can be applied simultaneously again. As Toffoli gates and controlled-single-qubit gates can be implemented with constant size, the overall depth, including the copying stage and the inverse stage, is $O(\log s+d)$, and the overall size is $O(s)$. $\hfill\square$

More basic concepts about quantum computation can be found in \cite{nielsen2010quantum}.

\subsection{Fan-in and fan-out gate}\label{subsec:fan-in-out}

A fan-in gate in quantum computation applies the following transformation.

\begin{equation}
\begin{aligned}
    &\ket{x_0}\ket{x_1x_2\ldots x_n} \to \ket{\bigoplus_{j=0}^n x_j}\ket{x_1x_2\dots x_n}, \\
    &\forall (x_0, x_1, \ldots, x_n)\in \mathbb{F}_2^{n+1}.
\end{aligned}
\end{equation}

On the other hand, a fan-out gate reverses the roles of the target and control qubits compared to a fan-in gate. Specifically,  a fan-out gate in quantum computation applies the following transformation.

\begin{equation}
\begin{aligned}
&\ket{x_0}\ket{x_1x_2\ldots x_n} \to \ket{x_0}\otimes(\bigotimes_{j=1}^n\ket{x_j\oplus x_0}), \\
&\forall (x_0, x_1, \ldots, x_n)\in \mathbb{F}_2^{n+1}.
\end{aligned}
\end{equation}

\begin{lemma}\label{Fan-in&out}
The $(n+1)$-qubit fan-in gate can be implemented by a CNOT circuit with $O(\log n)$-depth without ancillary qubits.
\end{lemma}


\noindent \textit{Proof.} Suppose a fan-in gate with $n$ control qubits can be applied in a depth of $f(n)$. Assume $n$ is an even number (for simplicity). In the first layer of the circuit, we apply $n/2$ CNOT gates, each targeting on $q_{2j-1}$ and controlled by $q_{2j}, j\in\{1, \ldots, n/2\}$. Thus, the value of $q_{2j-1}$ will become $x_{2j-1}\oplus x_{2j}$. If we continue applying a smaller fan-in gate on $q_0$ (as the target qubit) and $q_{2j-1}, j\in\{1, \ldots, n/2\}$ (as the control qubits),  the final value on $q_0$ will be $\bigoplus_{j=0}^n x_j$,  which is exactly what we desired. Next, we add a final layer,  which is the inverse of the first layer, so that all the $q_j, j\in\{1, \ldots, n\}$ will be restored. Hence, we get $f(n)=1+f(\frac{n}{2})+1=2+f(\frac{n}{2})$. This means $f(n) \in O(\log n)$. \rightline {$\hfill\square$}

The example of an $(8+1)$-qubit fan-in gate is shown in Fig.~\ref{fan-inex}.

\begin{figure*}
    \centering
   \begin{tikzpicture}
      \begin{yquant}[register/separation=2mm]
        qubit {$q_0$} q0;
        qubit {$q_1$} q1;
        qubit {$q_2$} q2;
        qubit {$q_3$} q3;
        qubit {$q_4$} q4;
        qubit {$q_5$} q5;
        qubit {$q_6$} q6;
        qubit {$q_7$} q7;
        qubit {$q_8$} q8;

        cnot q0|q1;
        cnot q0|q2;
        cnot q0|q3;
        cnot q0|q4;
        cnot q0|q5;
        cnot q0|q6;
        cnot q0|q7;
        cnot q0|q8;

        text {\space  = \space } (-);
        cnot q1|q2;
        cnot q3|q4;
        cnot q5|q6;
        cnot q7|q8;
        cnot q1|q3;
        cnot q5|q7;
        cnot q1|q5;
        cnot q0|q1;
        cnot q1|q5;
        cnot q5|q7;
        cnot q1|q3;
        cnot q7|q8;
        cnot q5|q6;
        cnot q3|q4;
        cnot q1|q2;

      \end{yquant}
   \end{tikzpicture}
    \caption{Two ways to implement a $(1+8)$-qubit fan-in gate.}
    \label{fan-inex}
\end{figure*}

For the fan-out gate,  note the equality in Fig.~\ref{basistrans}. Therefore,  by adding the $H$ gates on each end of each qubit,  the rest we need to do is just a fan-in gate (in fact,  we have to add $2n$ $H$ gates on $q_0$,  but two adjacent $H$ gates will be eliminated as $H^2=I$)
\begin{equation}
U_{\text{fan-out}}=H^{\otimes (n+1) }U_{ \text{fan-in} }H^{\otimes (n+1) }.
\end{equation}
Thus, an $(n+1)$-qubit fan-out gate can also be implemented by a circuit with $O(\log n)$-depth.

\begin{figure}[h]
    \centering
   \begin{tikzpicture}
      \begin{yquant}[register/separation=2mm]
        qubit {} q1;
        qubit {} q2;

        cnot q1|q2;
        text {\space  = \space } (-);
        box {$H$} q1, q2;
        cnot q2|q1;
        box {$H$} q1, q2;

      \end{yquant}
   \end{tikzpicture}
    \caption{CNOT basis transformation.}
    \label{basistrans}
\end{figure}

\subsection{Unary Encoding}

We will frequently use a special kind of encoding,  the Unary Encoding \cite{johri2021nearest}. The Unary Encoding can prepare the state
\begin{equation}
\sum_{i=1}^l \alpha_i\ket{e_i}
\end{equation}
for arbitrary $(\alpha_1, \alpha_2, \dots, \alpha_l)^T\in \mathbb{R}^l$ satisfying
$\sum_{i=1}^l |\alpha_i|^2=1$.
\begin{lemma} \label{Unary Encoding}
$\sum_{i=1}^l \alpha_i\ket{e_i}$ can be prepared by a quantum circuit with $O(\log{l})$-depth.
\end{lemma}
 \noindent \textit{Proof.} Here we briefly introduce the quantum gates used in the Unary Encoding. An $RBS(\theta)$ gate keeps the input $\ket{00}$ unchanged and turns the input $\ket{10}$ into $\cos{\theta}\ket{10}+\sin{\theta}\ket{01}$. Figure~\ref{fig:RBSdecomp} is an available way to construct the $RBS(\theta)$ gate.

The main idea of the Unary Encoding is to split the coefficient into different parts in every layer.
The parameter in every $RBS(\theta)$ gate is determined by the vector $\{\alpha_1, \alpha_2, \dots, \alpha_l\}$. We first apply an $X$ gate on $q_1$ so that the state becomes $\ket{e_1}$. Note that at this moment, $\ket{e_1}$ is holding the total amplitude. Then, we apply a $RBS(\theta)$ gate on $q_1$ and $q_{\lfloor l/2 \rfloor+1}$,  where $\theta$ is set as $\arccos{\sqrt{\alpha_1^2+\alpha_2^2+\cdots+\alpha_{\lfloor l/2 \rfloor}^2}}$,  and the state will become $\sqrt{\alpha_1^2+\alpha_2^2+\cdots+\alpha_{\lfloor l/2 \rfloor}^2}\ket{e_1}+\sqrt{\alpha_{\lfloor l/2 \rfloor+1}^2+\cdots+\alpha_l^2}\ket{e_{\lfloor l/2 \rfloor+1}}$. Now, $\ket{e_1}$ is holding the cumulative amplitude of $\ket{e_1}$,  $\ket{e_2}$, \dots $\ket{e_{\lfloor l/2 \rfloor}}$,  while $\ket{e_{\lfloor l/2 \rfloor+1}}$ is holding the cumulative amplitude of $\ket{e_{\lfloor l/2 \rfloor+1}}$, \dots $\ket{e_l}$. Thus, the $RBS(\theta)$ gate effectively splits the the total amplitude into two parts. Each part is yet another Unary Encoding on half of the qubits,  and the two parts can work in parallel. If we denote the depth of an $l$-qubit Unary Encoding as $f(l)$,  we get $f(l)=O(1)+f(\frac{l}{2})$. Hence $f(l)\in O(\log l)$.$\hfill\square$

Taking $l=8$ as an example, the Unary Encoding circuit is shown in Fig.~\ref{fig:Unary Example}. After the first $RBS(\theta)$ gate, the state becomes $\sqrt{\alpha_1^2+\alpha_2^2+\alpha_3^2+\alpha_4^2}\ket{e_1}+\sqrt{\alpha_5^2+\alpha_6^2+\alpha_7^2+\alpha_8^2}\ket{e_{5}}$, so that the later circuits on the top half and the bottom half can work at the same time. Notably, the same circuit structure can also be used to encode complex coefficients. By replacing $RBS(\theta)$ with a new two qubit gate that keeps the input $\ket{00}$ unchanged and turns the input $\ket{10}$ into $e^{i\alpha}\cos{\theta}\ket{10}+e^{i\beta}\sin{\theta}\ket{01}$, we can easily handle complex coefficients. This change does not affect the complexity.
\begin{figure}[h]
    \centering
    \begin{subfigure}[b]{0.45\textwidth}
    \centering
    \begin{tikzpicture}
\begin{yquant*}
qubit {} q1;
qubit {} q2;
[name=B] box {$B$} q1;
[name=S] box {$S$} q2;
\draw (B) -- (S);
text {$=$} (-);
cnot q1|q2;
box {$R_y(2\theta)$} q2 | q1;
cnot q1|q2;
\end{yquant*}
\end{tikzpicture}
\caption{The decomposition of the $RBS(\theta)$ gate.}
\label{fig:RBSdecomp}
    \end{subfigure}
\hfill
    \begin{subfigure}[b]{0.45\textwidth}
    \centering
   \begin{tikzpicture}
      \begin{yquant}[register/separation=2mm]
        qubit {$q_1$} q1;
        qubit {$q_2$} q2;
        qubit {$q_3$} q3;
        qubit {$q_4$} q4;
        qubit {$q_5$} q5;
        qubit {$q_6$} q6;
        qubit {$q_7$} q7;
        qubit {$q_8$} q8;

        box {$X$} q1;
        controlbox {$S$} q5;
        box {$B$} q1 | q5;
        controlbox {$S$} q3;
        box {$B$} q1 | q3;
        controlbox {$S$} q7;
        box {$B$} q5 | q7;
        controlbox {$S$} q2;
        box {$B$} q1 | q2;
        controlbox {$S$} q4;
        box {$B$} q3 | q4;
        controlbox {$S$} q6;
        box {$B$} q5 | q6;
        controlbox {$S$} q8;
        box {$B$} q7 | q8;
      \end{yquant}
   \end{tikzpicture}
    \caption{The Unary Encoding on 8 qubits, where the parameters in every $RBS(\theta)$ gate are determined by the vector $(\alpha_1,\alpha_2,\dots,\alpha_8)^T$.}
    \label{fig:Unary Example}
    \end{subfigure}
    \caption{The Unary Encoding technique.}
    \label{fig:Unary Encoding}
\end{figure}

\subsection{Hamming Weight Counter}

\begin{definition}
For $n\in \mathbb{N}^*, k\in \mathbb{N}$ and $k\leq n$, a Hamming Weight Counter (HWC) is an $n$-controlled quantum gate, where an $X$ gate is applied to the last qubit when the Hamming Weight of the first $n$ qubits is exactly $k$. It is formally defined by its action on computational basis states:

\begin{small}
\begin{equation}
\text{HWC}^{n+1}_k \ket{x_1 \ldots x_n}\ket{b} = 
\begin{cases} 
\ket{x_1 \ldots x_n}\ket{b \oplus 1} & \text{if } \sum_{i} x_i = k, \\
\ket{x_1 \ldots x_n}\ket{b} & \text{otherwise}.
\end{cases}
\nonumber
\end{equation}
\end{small}
\end{definition}
Essentially, an HWC gate checks if the Hamming Weight of the controlled qubits equals a required number and puts the result into the target qubit.

\begin{lemma}\label{lem:HWC}(\cite{zi2025shallow,rethinasamy2024logarithmic, zi2026efficient}) An $n$-controlled HWC gate can be implemented with $O((\log n)^2)$-depth and $O(n^2)$ size when using $O(\log k)$ ancillary qubits, or it can be implemented with $O(\log n)$-depth when using $o(n)$ ancillary qubits,
\end{lemma}

\section{Graph-structured states}\label{sec:k=2}
\subsection{General cases}
Before explaining how to prepare the most general HWP states, we will focus on a special case of $k=2$, called graph-structured states. While distinct from standard stabilizer graph states \cite{cao2023generation}, this nomenclature reflects their direct correspondence to the topology of undirected weighted graphs. The preparation of graph-structured states is fundamental and will play an important role in that of HWP states with $k \geq 3 $. Consider an undirected weighted graph $G=(V, E)$,  where $V=\{v_1, v_2, \dots, v_n\}$ denotes the set of vertices, $w_{ij}\in \mathbb{R}_+ \ (\forall~(v_i,v_j)\in E)$ indicates the weight of the corresponding edge and the number of edges is $m=|E|$. In the following text, we use $v_iv_j$ to denote graph edges instead of the notation $(v_i,v_j)$.‌ 
The graph-structured state is defined as
\begin{equation}\label{eq:graph-structured}
\ket{\psi_G}= \frac{1}{\sqrt{M}}\sum\limits_{i<j, v_iv_j\in E}w_{ij}\ket{e_i\oplus e_j},
\end{equation}
where $M=\sum_{i<j, v_iv_j\in E}w_{ij}^2$ is the normalization factor and $\ket{e_i}$ means the computational basis state that has an $1$ on the $i^{th}$ qubit and $0$s on all other qubits. In this mapping, each edge $v_i v_j$ in the graph corresponds to a basis state $\ket{e_i\oplus e_j}$ in the two-excitation manifold. As an illustrative example, for the graph structure shown in Fig.~\ref{Graph Example}, the target state is expressed as
\begin{equation}\label{eq}
\begin{aligned}
 \frac{1}{\sqrt{18}} \Big( \sqrt{2}\ket{1100000}+\sqrt{3}\ket{0110000} + \sqrt{7}\ket{0011000}+ \\ \sqrt{3}\ket{0100100} + \sqrt{2}\ket{0000110}+\sqrt{1}\ket{0000101} \Big).
\end{aligned}
\end{equation}

\begin{figure}[h]
    \centering
    \begin{tikzpicture}
    \node[circle,  draw,  fill=black,  inner sep=2pt] (v1) at (0, 0) {};
    \node[circle,  draw,  fill=black,  inner sep=2pt] (v2) at (1, -1) {};
    \node[circle,  draw,  fill=black,  inner sep=2pt] (v3) at (0, -2) {};
    \node[circle,  draw,  fill=black,  inner sep=2pt] (v4) at (1, -3) {};
    \node[circle,  draw,  fill=black,  inner sep=2pt] (v5) at (2, -1) {};
    \node[circle,  draw,  fill=black,  inner sep=2pt] (v6) at (3, -3) {};
    \node[circle,  draw,  fill=black,  inner sep=2pt] (v7) at (3, 0) {};

    \node[above] at (v1) {$v_1$};
    \node[left] at (v2) {$v_2$};
    \node[left] at (v3) {$v_3$};
    \node[below] at (v4) {$v_4$};
    \node[right] at (v5) {$v_5$};
    \node[below] at (v6) {$v_6$};
    \node[above] at (v7) {$v_7$};

    \draw (v1) -- node[above] {$\sqrt{2}$} (v2);
    \draw (v2) -- node[above] {$\sqrt{3}$} (v5);
    \draw (v2) -- node[left] {$\sqrt{3}$} (v3);
    \draw (v5) -- node[left] {$\sqrt{1}$} (v7);
    \draw (v4) -- node[left, below] {$\sqrt{7}$} (v3);
    \draw (v5) -- node[left] {$\sqrt{2}$} (v6);

    \end{tikzpicture}

    \caption{An undirected weighted graph whose corresponding state is shown in Eq.~(\ref{eq}).}
    \label{Graph Example}
\end{figure}

We first introduce $l=m$ (recall that $m$ is the number of edges, $m=|E|$) ancillary qubits $\{q_1', q_2', \dots, q_{m}'\}$, where each $q_j'$ corresponds to the edge $e_j\in E$. Meanwhile,  $\ket{e_j'}$ has the same meaning as $\ket{e_j}$,  but we use it only to describe states on ancillary qubits. Denote the two vertices of the edge $e_j$ as $v_{j_1}$ and $v_{j_2}$. The detailed QSP procedure for a general graph-structured state is shown in Algorithm \ref{Arbal}.

\begin{breakablealgorithm}
    \caption{Preparation of a general graph-structured quantum state with $O(m)$ ancillary qubits.}
    \label{Arbal}
    \begin{algorithmic}[1]
    \Require An arbitrary graph $G=(V, E)$,  where $n=|V|$; $n$ working qubits together with $m$ ancillary qubits.
    \Ensure The circuit that prepares $\ket{\psi_G}\ket{0}^{\otimes m}$.
    
    \State Apply Unary Encoding on ancillary qubits,  where the coefficient of $\ket{e_j'}$ is the weight of $e_j$ divided by $\sqrt{M}$.

    \For{every edge $e_j=v_{j_1}v_{j_2}$}
        \State Apply 2 CNOT gates both controlled by $q_j'$,  and targeting on $q_{j_1}$, $q_{j_2}$ respectively.
    \EndFor

    \For{every edge $e_j=v_{j_1}v_{j_2}$}
        \State Apply a Toffoli gate controlled by $q_{j_1}$ and $q_{j_2}$,  targeting on $q_j'$.
    \EndFor
    \State Return the whole circuit.
    \end{algorithmic}
\end{breakablealgorithm}

\begin{lemma}
    Algorithm \ref{Arbal} can correctly prepare the general graph-structured quantum state.
\end{lemma}

\noindent \textit{Proof.} By dividing the whole algorithm into 3 steps: line 1; line 2 to line 4; line 5 to line 7, we can easily check the state transformation after each step,

\begin{equation}
\begin{aligned}
    & \ket{0}^{\otimes n}\otimes\ket{0}^{\otimes m} \\
     \xrightarrow{\text{~~1~~~}} & \frac{1}{\sqrt{M}}\sum\limits_{e_j=v_{j_1}v_{j_2}\in E}w_{j_1j_2}\ket{0}^{\otimes n}\otimes\ket{e_j'} \\
     \xrightarrow{\text{2 to 4}} & \frac{1}{\sqrt{M}}\sum\limits_{e_j=v_{j_1}v_{j_2}\in E}w_{j_1j_2}\ket{e_{j_1}\oplus e_{j_2}}\otimes\ket{e_j'} \\
     \xrightarrow{\text{5 to 7}} & \frac{1}{\sqrt{M}}\sum\limits_{e_j=v_{j_1}v_{j_2}\in E}w_{j_1j_2}\ket{e_{j_1}\oplus e_{j_2}}\otimes\ket{0}^{\otimes n}.
\end{aligned}
\end{equation}
Thus, the algorithm is correct. For the case of complex coefficients, we simply encode them in the Unary Encoding stage. $\hfill\square$

\begin{theorem} \label{thm:Graph Arbitrary}
    For an arbitrary graph $G$ with $n$ vertices,  $\ket{\psi_G}$ can be prepared by a circuit with $O(\log n)$-depth using $O(m)$ ancillary qubits.
\end{theorem}

\noindent \textit{Proof.} The first step of Algorithm \ref{Arbal} (line 1) is the Unary Encoding circuit on $m$ ancillary qubits. Hence by Lemma \ref{Unary Encoding},  step 1 can be implemented in a depth of $O(\log m)$. As $m<\frac{n^2}{2}$,  $O(\log m)=O(\log n)$.

For the second step (line 2 to 4),  we will totally apply $2m$ CNOT gates. These gates are obviously commutable. We can divide the $2m$ CNOT gates into 2 sets,  each with $m$ CNOT gates. For $j\in\{1, 2, \dots, m\}$,  we put one CNOT gate controlled by $q_j'$ into the first set, and put the other one into the second set. We can find that the CNOT gates in a single set can be treated as several independent fan-in gates. By Lemma \ref{Fan-in&out},  applying the CNOT gates in a single set can be implemented in a depth of $O(\log n)$. As there are only 2 sets,  the $2m$ CNOT gates can be implemented in a depth of $O(\log n)+O(\log n)=O(\log n)$.

For the final step (line 5 to 7), we apply $m$ Toffoli gates. Every $q_j', j\in\{1, 2, 3, \dots, m\}$ will be used as target qubit exactly once,  while some $q_i$ will be used as control qubit for more than once,  suppose $t_i$ times. We can optimize the depth using the following method.
We first apply a copying stage. We copy each $q_i$ for $t_i-1$ times (by adding $t_i$ more ancillary qubits). Note that the number of ancillary qubits that are newly added in this step is $\sum_{i=1}^n (t_i-1)=2m-n=O(m)$,  and the copying stage has a depth of $\log{(\max_i{t_i})}=O(\log n)$. Secondly,  we apply $m$ Toffoli gates by using $q_i$ or its copy as control qubits. As all Toffoli gates run independently,  the depth of this stage is just the depth of one Toffoli gate,  which is a constant. Finally, we apply the inverse operation of the copying stage. Thus, the final step can be implemented in a depth of $O(\log n)+O(\log n)=O(\log n)$.

Combining all three steps together, we can get that $\ket{\psi_G}$ can be generated in a depth of $O(\log n)$,  when using $O(m)$ ancillary qubits.$\hfill\square$

To illustrate Algorithm~\ref{Arbal}, consider the input graph $G$ shown in Fig.~\ref{Graph Example}. The output is the QSP circuit for $\ket {\psi_G}$ in Fig.~\ref{example}. For the first step (line 1), we apply the Unary Encoding circuit, the parameters in all $RBS$ gates depend on the arbitrary weights. For the second step (line 2 to 4) we apply $2\times 6=12$ CNOT gates, and these CNOT gates can be divided into two parts (denoted as normal and bold style). CNOT gates in the same style belong to the same part, and can be applied in $O(\log 6)$-depth. For the final step (line 5 to 7), we apply 6 Toffoli gates. We have completed the preparation of $\ket {\psi_G} $.

\begin{figure}
    \centering
\scalebox{0.75}{
\begin{tikzpicture}
    \node[] (v1) at (0.45, 0.1) {};
    \node[above] at (v1) {(\romannumeral1)};
    \node[] (v2) at (4.45, 0.1) {};
    \node[above] at (v2) {(\romannumeral2)};
    \node[] (v3) at (9, 0.1) {};
    \node[above] at (v3) {(\romannumeral3)};
\begin{yquant*}[operator/separation=0.02mm]
qubit {$q_1$} a;
qubit {$q_2$} a[+1];
qubit {$q_3$} a[+1];
qubit {$q_4$} a[+1];
qubit {$q_5$} a[+1];
qubit {$q_6$} a[+1];
qubit {$q_7$} a[+1];
qubit {$q_1'$} b;
qubit {$q_2'$} b[+1];
qubit {$q_3'$} b[+1];
qubit {$q_4'$} b[+1];
qubit {$q_5'$} b[+1];
qubit {$q_6'$} b[+1];
box {UE} (b);
barrier a, b;

[operator style={very thick},control style={very thick}]cnot a[0] | b[0];
cnot a[1] | b[0];

[operator style={very thick},control style={very thick}]cnot a[1] | b[1];
cnot a[4] | b[1];

[operator style={very thick},control style={very thick}]cnot a[4] | b[2];
cnot a[6] | b[2];

[operator style={very thick},control style={very thick}]cnot a[1] | b[3];
cnot a[2] | b[3];

[operator style={very thick},control style={very thick}]cnot a[2] | b[4];
cnot a[3] | b[4];

[operator style={very thick},control style={very thick}]cnot a[4] | b[5];
cnot a[5] | b[5];

barrier a, b;

cnot b[0] | a[0, 1];
cnot b[1] | a[1, 4];
cnot b[2] | a[4, 6];
cnot b[3] | a[1, 2];
cnot b[4] | a[2, 3];
cnot b[5] | a[4, 5];

\end{yquant*}
\end{tikzpicture}
}
    \caption{The circuit for preparing $\ket{\psi_G}$,  where $G$ is shown in Fig.~\ref{Graph Example}. In the middle stage,  CNOT gates of the same style (normal or bold) can be applied in depth $O(\log m)$.}
    \label{example}
\end{figure}

While the general graph preparation scheme is asymptotically optimal, the $O(m)$ ancillary qubits overhead poses a challenge for near-term quantum hardware. More importantly, investigating states with specific graph topologies is motivated by their profound physical significance. For instance, quantum states associated with tree structures and grid structures are intimately connected to the eigenstates of integrable systems, such as the Gaudin model \cite{gaudin1976diagonalisation, dukelsky2004colloquium,bortz2007exact} (also known as the central spin model). To bridge the gap between theoretical optimality and physical applicability, we further demonstrate that for such structured graphs, the ancillary qubits can be entirely eliminated without compromising the $O(\log n)$ circuit depth. This results in an ancilla-free, hardware-efficient preparation strategy for a broad class of physically relevant quantum states.

\subsection{Tree-structured states}

We introduce several additional methods for optimizing certain CNOT circuits.

It is well known that any $n$-qubit CNOT circuit can be viewed as an invertible linear map over $\mathbb{F}_2^n$,  hence it can be represented by an invertible matrix in $\mathbb{F}_2^{n \times n}$.

A chain-like CNOT circuit consists of a sequence of CNOT gates applied to $q_1,  q_2,  \ldots,  q_n$,  and the $j^{th}$ CNOT gate is controlled by $q_j$ and targets on $q_{j+1}$. Taking $n=5$ as an example,  its circuit and matrix representation are shown below

\begin{figure}[h]
    \centering
    \begin{subfigure}[c]{0.30\textwidth}
        \centering
   \begin{tikzpicture}
      \begin{yquant}[register/separation=2mm]
        qubit {$q_1$} q1;
        qubit {$q_2$} q2;
        qubit {$q_3$} q3;
        qubit {$q_4$} q4;
        qubit {$q_5$} q5;
        cnot q2|q1;
        cnot q3|q2;
        cnot q4|q3;
        cnot q5|q4;

      \end{yquant}
   \end{tikzpicture}
    \end{subfigure}
    \begin{subfigure}[c]{0.30\textwidth}
        \centering
        \begin{equation*}
A=
\begin{pmatrix}
1 &   &   &   & \\
1 & 1 &   &   & \\
1 & 1 & 1 &   & \\
1 & 1 & 1 & 1 & \\
1 & 1 & 1 & 1 & 1
\end{pmatrix}
, A^{-1}=
\begin{pmatrix}
1 &   &   &   & \\
1 & 1 &   &   & \\
0 & 1 & 1 &   & \\
0 & 0 & 1 & 1 & \\
0 & 0 & 0 & 1 & 1
\end{pmatrix}.  
\end{equation*}
    \end{subfigure}
    \caption{A chain-like CNOT circuit on 5 qubits and its matrix representation.}
    \label{chainlike}
\end{figure}

\begin{lemma} \label{Chain}
(\cite{jiang2020optimal}) The $n$-qubit chain-like CNOT circuit can be efficiently implemented by a CNOT circuit of $O(\log n)$-depth without ancillary qubits.  
\end{lemma}

\begin{figure}[h]
    \centering
    \begin{subfigure}[c]{0.45\textwidth}
    \centering
       \begin{tikzpicture}
      \begin{yquant}[register/separation=2mm]
        qubit {$q_{a1}$} qa1;
        qubit {$q_{a2}$} qa2;
        qubit {$q_{a3}$} qa3;
        qubit {$q_{a4}$} qa4;
        qubit {$q_{b1}$} qb1;
        qubit {$q_{b2}$} qb2;
        qubit {$q_{b3}$} qb3;

        cnot qb1|qa1;
        cnot qb2|qa1;
        cnot qb2|qa2;
        cnot qb3|qa2;
        cnot qb1|qa3;
        cnot qb2|qa4;

        \end{yquant}
    \end{tikzpicture}
    \end{subfigure}
    \hfill
    \begin{subfigure}[c]{0.45\textwidth}
    \centering
    \begin{tikzpicture}
    \node[circle,  draw,  fill=black,  inner sep=2pt] (a1) at (0, 0) {};
    \node[circle,  draw,  fill=black,  inner sep=2pt] (a2) at (0, -1) {};
    \node[circle,  draw,  fill=black,  inner sep=2pt] (a3) at (0, -2) {};
    \node[circle,  draw,  fill=black,  inner sep=2pt] (a4) at (0, -3) {};
    \node[circle,  draw,  fill=black,  inner sep=2pt] (b1) at (2, -0.5) {};
    \node[circle,  draw,  fill=black,  inner sep=2pt] (b2) at (2, -1.5) {};
    \node[circle,  draw,  fill=black,  inner sep=2pt] (b3) at (2, -2.5) {};

    \node[left] at (a1) {$q_{a1}$};
    \node[left] at (a2) {$q_{a2}$};
    \node[left] at (a3) {$q_{a3}$};
    \node[left] at (a4) {$q_{a4}$};
    \node[right] at (b1) {$q_{b1}$};
    \node[right] at (b2) {$q_{b2}$};
    \node[right] at (b3) {$q_{b3}$};

    \draw[->] (a1) --  (b1);
    \draw[->] (a1) --  (b2);
    \draw[->] (a2) --  (b2);
    \draw[->] (a2) --  (b3);
    \draw[->] (a3) --  (b1);
    \draw[->] (a4) --  (b2);    
    \end{tikzpicture}
    \end{subfigure}

    \caption{A CNOT circuit and its corresponding bipartite graph.}
    \label{biex}
\end{figure}
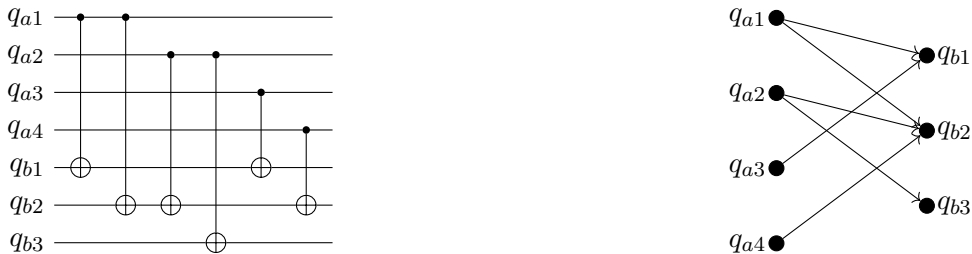

One next approach is for bipartite CNOT circuits. The qubits in a bipartite CNOT circuit can be divided into two parts $A=\{q_{a_1}, q_{a_2}, \ldots, q_{a_n}\}$ and $B=\{q_{b_1}, q_{b_2}, \ldots, q_{b_{n'}}\}$, and every CNOT gate must be controlled by a qubit in $A$ and target on a qubit in $B$. 
Thus,  this kind of CNOT circuits can be described by the corresponding directed bipartite graph $(A\sqcup B, E)$,  where $q_{a_i}q_{b_j}\in E$ iff there is a CNOT gate controlled by $q_{a_i}$ and targeting on $q_{b_j}$ in the circuit. Again, we set $m=|E|$. We can also give a matrix representation of a bipartite CNOT circuit. Let $\vec{u}_A=(x_{a_1}, x_{a_2}, \ldots, x_{a_n})\in \mathbb{F}_2^{n}$ be some computational basis on qubits in $A$. Similarly, let $\vec{u}_B=(x_{b_1}, x_{b_2}, \ldots, x_{b_{n'}})\in \mathbb{F}_2^{n'}$ be some computational basis on qubits in $B$. Further let $M\in \mathbb{F}_2^{n' \times n}$, then $(\vec{u}_A, \vec{u}_B)\to(\vec{u}_A, \vec{u}_B+M\vec{u}_A)$ can be seen as a CNOT circuit based on a directed bipartite graph.
Figure~\ref{biex} is an example of a bipartite CNOT circuit with 6 CNOT gates.

It is obvious that all the CNOT gates in the circuit commute with each other, which means that the order of those CNOT gates does not matter. Suppose we are provided with $l$ dirty ancillary qubits---where a ``dirty'' qubit refers to one whose initial state is unknown but must be restored to its original state by the end of the operation---and that the circuit contains $m$ CNOT gates in total.

\begin{lemma} \label{CNOTbipartite}
When using $l$ dirty ancillary qubits,  any bipartite CNOT circuit with $m$ edges can be implemented by a circuit with $O(\frac{m}{l}\log l)$-depth and $O(m)$-size.
\end{lemma}

\noindent \textit{Proof.} Since the CNOT gates mutually commute,  we plan to apply $l$ CNOT gates from the $m$ CNOT gates in depth $O(\log l)$ (the $l$ CNOT gates can be chosen in any way),  so that the entire circuit can be implemented in a depth of $O(\frac{m}{l}\log l)$.

We name the $l$ CNOT gates as $S=\{CX_1, CX_2, \ldots, CX_l\}$ (note that $S$ can also be treated as an operator by applying all the gates inside $S$),  and name the $l$ dirty ancillary qubits $\{q_1', q_2', \ldots, q_l'\}$. For $CX_j$,  which is controlled by some $q_a$ and targets on some $q_b$,  we turn it into 2 CNOT gates. One is controlled by $q_a$ and targets on $q_j'$,  named as $CX_j^a$,  and the other one is controlled by $q_j'$ and targets on $q_b$,  named as $CX_j^b$. It is obvious that the original $CX_j$ is not equivalent to $CX_j^A$ followed by $CX_j^B$ when $q_j$ is dirty,  so we have to do something more complicated. 

We denote the set of all $CX_j^a$ as $S_A$,  the set of all $CX_j^b$ as $S_B$. Again, all the CNOT gates in $S_A$ commute with each other, so do the gates in $S_B$. Thus, we can apply all gates in $S_A$($S_B$) without concerning the order of applying those CNOT gates. We further notice that $S_A$ consists of a group of independent fan-in gates,  while $S_B$ consists of a group of independent fan-out gates,  and every single fan-in (out) gate concerns less than $n+1$ qubits. This means that we can apply $S_A$ and $S_B$ both in a depth of $O(\log l)$. To synthesize the operation $S$ using $S_A$ and $S_B$, we employ the sequence $S=S_AS_BS_AS_B$. The proof for the correctness of the former formula is not difficult either. Let us just consider the value on $q_a, q_j', q_b$ step by step when applying 
$S_AS_BS_AS_B$,  and check if the final state coincides with $CX_j$.

\begin{equation}
\begin{aligned}
     & \ket{x_a}\ket{x_j'}\ket{x_b} \\
     \xrightarrow{S_B} & \ket{x_a}\ket{x_j'}\ket{x_b\oplus x_j'} \\
     \xrightarrow{S_A} & \ket{x_a}\ket{x_j'\oplus x_a}\ket{x_b\oplus x_j'} \\
     \xrightarrow{S_B} & \ket{x_a}\ket{x_j'\oplus x_a}\ket{x_b\oplus x_a} \\
     \xrightarrow{S_A} & \ket{x_a}\ket{x_j'}\ket{x_b\oplus x_a}.
\end{aligned}
\end{equation}

The final state exactly matches the desired output of $CX_j$, while leaving the dirty ancillary qubit restored to its original state. Hence,  we can apply $S$ in $O(\log l)+O(\log l)+O(\log l)+O(\log l)=O(\log l)$-depth and $O(l)$-size. Thus the bipartite CNOT circuit can be implemented in $O(\frac{m}{l}\log l)$-depth and $O(\frac{m}{l}l)=O(m)$-size. $\hfill\square$

Obviously,  the larger $l$ in Lemma \ref{CNOTbipartite} is,  the more significantly the circuit depth can be reduced. As a result,  we will employ as many ancillary qubits as possible in later operations.

The last approach is for general CNOT circuits. Here we prove that any CNOT circuit on $n$ qubits can be implemented in a depth of $O(\log n)$ with $O(n^2)$ dirty ancillary qubits.

\begin{lemma}\label{CNOTgeneral}
When using $O(n^2)$ dirty ancillary qubits, if the matrix representation of a CNOT circuit can be reduced to Jordan canonical form, then this CNOT circuit on $n$ qubits can be implemented by a circuit with $O(\log n)$-depth.
\end{lemma}

\noindent \textit{Proof.} Assume the matrix representation of a general CNOT circuit is $M\in \mathbb{F}_2^{n \times n}$. Given that $M$ is invertible,  $M$ could be expressed as $M=PJP^{-1}$,  where $P$ is invertible and $J$ is the Jordan matrix whose diagonal elements are all 1.

Given an input $\vec{x}=(x_1, x_2, \ldots, x_n)\in \mathbb{F}_2^n$,  we wish the output could be $M\vec{x}$. We first take $n$ dirty ancillary qubits,  and suppose the value on these $n$ ancillary qubits is $\vec{y}=(y_1, y_2, \ldots, y_n)\in \mathbb{F}_2^n$.

We can solve this problem by the following steps.

\begin{equation}
\begin{aligned}
     & (\vec{x}, \vec{y})\\
     \xrightarrow{~~~~~1~~~~~} & (\vec{x}, \vec{y}+P^{-1}\vec{x})\\
     \xrightarrow{~~~~~2~~~~~} & (\vec{x}+P(\vec{y}+P^{-1}\vec{x}), \vec{y}+P^{-1}\vec{x})=(P\vec{y}, \vec{y}+P^{-1}\vec{x})\\
     \xrightarrow{~~~~~3~~~~~} & (P\vec{y}, \vec{y}+P^{-1}\vec{x}+P^{-1}(P\vec{y}))=(P\vec{y}, P^{-1}\vec{x})\\
     \xrightarrow{~~~~~4~~~~~} & (P\vec{y}, JP^{-1}\vec{x})\\
     \xrightarrow{\text{repeat } 1, 2, 3} & (PJP^{-1}\vec{x}, \vec{y})=(M\vec{x}, \vec{y}).
\end{aligned}
\end{equation}

Here,  step 1, 2, 3 are bipartite CNOT circuits. In step 1 and 3,  working qubits are control qubits and ancillary qubits are target qubits,  and they shift the roles in step 2. As there are at most $n \times n=n^2$ edges in each bipartite graph,  if we introduce $O(n^2)$ more ancillary qubits,  step 1, 2, 3 can be implemented in a depth of $O(\frac{n^2}{n^2}\log{n^2})=O(\log n)$. Step 4 consists of several independent CNOT circuits, each corresponding to the inverse of a chain-like CNOT circuit, so the depth of step 4 is also $O(\log n)$. Therefore the overall depth is $O(\log n)$.$\hfill\square$

Before we present the preparation algorithm,  it is necessary to review a significant property of trees.
\begin{lemma} \label{lem}
(\cite{Diestel2017}) For every tree $G$ with $n$ vertices where $n\ge3$,
there exists a vertex $v$,  such that every connected component in $G-v$ has fewer than $(n-1)/2$ edges and no more than $n/2$ vertices.
\end{lemma}

We can get a corollary from this lemma.
\begin{corollary} \label{cor}
For every tree $G=(V, E)$ with $n$ vertices and $r<n$,  there exists a vertex set $V'\subset V$,  such that $|V'| \le 2r$ and every connected component in $G-V'$ has no more than $n/r$ vertices.
\end{corollary}

\noindent \textit{Proof.} Here we only prove when $r=2^t$ is a power of 2. We prove it by induction on $t$.

When $t=1$,  the claim is correct using the lemma above directly.
Assume the claim is correct when $r=2^s$. For $t=s+1$,  by induction,  we already have a vertex set $V''\subset V$,  such that $|V''| \le 2\cdot2^s$ and every connected component in $G-V''$ has no more than $n/2^s$ vertices. But there may exist connected component in $G-V''$ that has more than $n/2^{s+1}$ vertices. The number of this kind of connected component will be no more than $2^{s+1}$. By Lemma \ref{lem},  we can add the central vertex of this kind of connected component in to $V''$ and hence get a new vertex set $V'$,  obviously every connected component in $G-V'$ has no more than $n/2^{s+1}$ vertices now. Besides, $|V'|\le |V''|+2^{s+1} \le 2\cdot2^{s+1}$, which completes the proof. $\hfill\square$

Now we are ready to show the circuit that prepares a tree-structured state in depth $O(\log n)$ without ancillary qubits.

\begin{breakablealgorithm}
    \caption{Preparation of tree-structured states without ancillary qubits.}
    \begin{algorithmic}[1]
    \Require An arbitrary tree $G=(V, E)$,  where $n=|V|$; $n$ working qubits.
    \Ensure The circuit that prepares $\ket{\psi_G}$.

    \State Choose $v_1$ as the root of $G$,  then every edge in $G$ has a parent node and a child node,  and every node has a height $h$. Denote the height of $G$ as $H$.
    \State Apply Unary Encoding on $\{q_2, q_3, \ldots, q_n\}$,  and note the qubit $q_j$ corresponds to the node $v_j$. The coefficient of $\ket{e_j}$ is (the weight of the edge of which $v_j$ acts as the child node)/$\sqrt{M}$.
    \For {$h=1$ to $H$}
        \State For every node $v_j$ which height is $h$,  apply a CNOT gate controlled by $v_j$ and targeting on its parent node.
    \EndFor
    \State Return the whole circuit.
    \end{algorithmic}
    \label{Treeal}
\end{breakablealgorithm}

\begin{lemma}
    Algorithm \ref{Treeal} can correctly prepare a tree-structured state.
\end{lemma}

\noindent \textit{Proof.} After line 1 and 2,  the state becomes 

\begin{equation}
\frac{1}{\sqrt{M}}\sum\limits_{v_iv_j\in E,  v_j \text{ is the child node of this edge}}w_{ij}\ket{e_j}.
\end{equation}
Except for the root node ($v_1$),  every node will act as the child node exactly once,  so the state above is well defined.

For every CNOT gate applied between line 3 to 5,  take the CNOT gate controlled by $v_j$ and targeting on its parent node,  $v_i$ as an example,  the only change that this CNOT gate makes is that it turns $\ket{e_j}$ into $\ket{e_i\oplus e_j}$. Thus, the whole circuit between line 3 to 5 successfully turns all $\ket{e_j}$ into $\ket{e_i\oplus e_j}$. Hence the state becomes
\begin{equation}
\frac{1}{\sqrt{M}}\sum\limits_{v_iv_j\in E,  v_j \text{ is the child node of this edge}}w_{ij}\ket{e_i\oplus e_j},
\end{equation}
which is exactly what we want. Also for the case of complex coefficients, we simply encode them in the Unary Encoding stage.$\hfill\square$

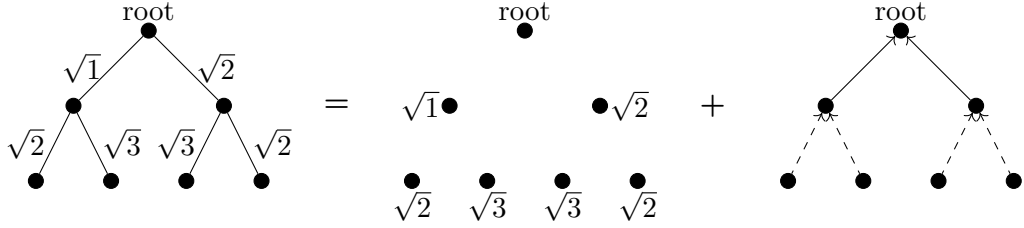
\begin{figure}
    \centering
    \begin{tikzpicture}  
        \node [circle,  draw,  fill=black,  inner sep=2pt](v4) at (0, 0) {};
        \node [circle,  draw,  fill=black,  inner sep=2pt](v5) at (1, 0) {};
        \node [circle,  draw,  fill=black,  inner sep=2pt](v6) at (2, 0) {};
        \node [circle,  draw,  fill=black,  inner sep=2pt](v7) at (3, 0) {};
        \node [circle,  draw,  fill=black,  inner sep=2pt](v2) at (0.5, 1) {};
        \node [circle,  draw,  fill=black,  inner sep=2pt](v3) at (2.5, 1) {};
        \node [circle,  draw,  fill=black,  inner sep=2pt](v1) at (1.5, 2) {};
        \node[above] at (v1) {root};
        \node[left] at (v2) {$\sqrt{1}$};
        \node[right] at (v3) {$\sqrt{2}$};
        \node[below] at (v4) {$\sqrt{2}$};
        \node[below] at (v5) {$\sqrt{3}$};
        \node[below] at (v6) {$\sqrt{3}$};
        \node[below] at (v7) {$\sqrt{2}$};
        \node[] () at (4, 1) {$\rightarrow$};

        \begin{scope}[xshift=5cm, on grid]
        \node [circle,  draw,  fill=black,  inner sep=2pt](v4) at (0, 0) {};
        \node [circle,  draw,  fill=black,  inner sep=2pt](v5) at (1, 0) {};
        \node [circle,  draw,  fill=black,  inner sep=2pt](v6) at (2, 0) {};
        \node [circle,  draw,  fill=black,  inner sep=2pt](v7) at (3, 0) {};
        \node [circle,  draw,  fill=black,  inner sep=2pt](v2) at (0.5, 1) {};
        \node [circle,  draw,  fill=black,  inner sep=2pt](v3) at (2.5, 1) {};
        \node [circle,  draw,  fill=black,  inner sep=2pt](v1) at (1.5, 2) {};
        \node[above] at (v1) {root};

        \draw[<-] (v1)--(v2);
        \draw[<-] (v1)--(v3);
        \draw[dashed, <-] (v2)--(v4);
        \draw[dashed, <-] (v2)--(v5);
        \draw[dashed, <-] (v3)--(v6);
        \draw[dashed, <-] (v3)--(v7);
        
        \end{scope}
    \end{tikzpicture} 
    \caption{Algorithm for preparing tree-structured states,  and nodes are labelled in level order (although the order does not really matter,  we set this order just to match the state described in the text).}
    \label{fig:treeex}
\end{figure}

An example is shown in Fig.~\ref{fig:treeex}. In this example,  we aim to prepare the state $\ket{\psi_G} = \frac{1}{\sqrt{13}} ( \sqrt{1} \ket{1100000} + \sqrt{2} \ket{1010000}+\sqrt{2} \ket{0101000}+\sqrt{3} \ket{0100100}+\sqrt{3} \ket{0010010}+\sqrt{2} \ket{0010001})$. The first block on the right-hand side of Figure~\ref{fig:treeex} means the Unary Encoding circuit. The coefficient of a non-root node is set to the weight of the edge in which the node acts as the child node. The second block means the CNOT gates circuits followed by the Unary Encoding circuit, the arrow means the CNOT gate controlled by the start of the arrow and targeting on the end of the arrow. The order of these CNOT gates matters. CNOT gates shown in solid arrows should be applied first, and the dashed arrows follow.

\begin{theorem}
     For an arbitrary tree $G$ with $n$ vertices,  $\ket{\psi_G}$ can be prepared by a circuit with $O(\log n)$-depth without ancillary qubits.
\end{theorem}

\noindent \textit{Proof.} Line 1 and 2 is the Unary Encoding circuit,  which can be implemented in a depth of $O(\log n)$.

For the circuit between line 3 to 5,  we apply a tree-shape CNOT circuit. For every edge in $G$,  we apply a CNOT gate controlled by child node and targeting on parent node.  But here the order matters. We have to apply these CNOT gates from top to bottom.

Here we consider the inverse of this circuit,  and call it $U$. This corresponds to a circuit that applies these CNOT gates from bottom to top. Note that after applying $U$,  the value on a single node is the sum of all the original value of its descendent and its original value.

We solve this problem by recursion. By using Corollary \ref{cor},  we can find a vertex set $V'\subset V$,  such that $|V'| \le 2r$ and every connected component in $G-V'$ has no more than $n/r$ vertices. If we correctly set all the values of vertices in $V'$ and further apply all the CNOT gates controlled by vertex in $V'$,  then we successfully delete these vertices from the whole circuit,  and for the rest of the CNOT circuit,  we can implement all the connected components independently. The process of deleting these vertices can be divided into 3 steps.

Step 1: For $\forall v_j\in V'$ define $V_j=\{v_i:v_i$ is a descendant of $v_j$ and the inner vertices in the path from $v_j$ to $v_i$ do not have vertices in $V'\}$,  and then apply a fan-in gate from $V_j$ to $v_j$. It is obvious that all the $V_j$ are independent from each other. So the whole depth of step 1 is $O(\log n)$.

Step 2: Make a new CNOT circuit inside $V'$: there is a CNOT gate controlled by $v_i\in V'$ and targeting on $v_j\in V'$ iff $v_i$ is a descendant of $v_j$ and the inner vertices in the path from $v_j$ to $v_i$ do not have vertices in $V'$. The sequence is also from bottom to top, hence the matrix representation of this CNOT circuit is a upper triangular matrix, which can be reduced to Jordan canonical form. We can find that after this step,  all of $v_j \in V'$ has possessed the correct value. Further more, the new CNOT circuit created in this step is applied to $2r$ vertices. Crucially,  based on Corollary \ref{cor},  when we set $r=\sqrt{n}$,  all the other qubits could be used as dirty ancillary qubit. We can take out $O((\sqrt{n})^2)=O(n)$ vertices as dirty ancillary qubits. By Lemma \ref{CNOTgeneral}, the new CNOT circuit can be implemented in a depth of $O(\log n)$.

Step 3: For $\forall v_j\in V'$,  apply a CNOT gate controlled by $v_j$ and targeting on the parent node of $v_j$ if the parent node is not in $V'$. This is a group of independent fan-in gates,  so step 3 can be implemented in a depth of $O(\log n)$.

As a result, the entire process can be implemented in a depth of $O(\log n)$.

If we denote the depth of the tree-shape CNOT circuit as $f(n)$,  by setting $r=\sqrt{n}$,  we can get $f(n)=O(\log n)+f(\sqrt{n})$. Hence $f(n)=O(\log n)$. Together with the former Unary Encoding,  we can claim that tree-structured states can be prepared in a depth of $O(\log n)+O(\log n)=O(\log n)$. $\hfill\square$

\subsection{Grid-structured states}

Compared to the vertex in the tree graph with explicit layer structure (parent and child,  ancestor and descendant),  the nodes in the grid graph have no similar relations, which makes it difficult to implement appropriate Unary Encoding directly. Therefore, we consider doing some pretreatment and applying the inverse circuit. This kind of method has appeared in other QSP problems \cite{gleinig2021efficient}.

Before introducing the algorithm,  we first consider the effect of a controlled-$RBS(\theta)$ gate. Assume that $q_1$ is controlling an $RBS(\theta)$ gate on $q_2, q_3$. 
When we just consider the computational basis with Hamming weight 2, besides a trivial input $\ket{0}^{\otimes n}$,  we can find that the controlled-$RBS(\theta)$ gate will only affect $\ket{e_1\oplus e_2}$ and $\ket{e_1\oplus e_3}$,  while keeping all other possible inputs unchanged. $\ket{e_1\oplus e_2}$ will turn into $\cos\theta\ket{e_1\oplus e_2}+\sin\theta\ket{e_1\oplus e_3}$. As a result,  when the input of the circuit is a state with graph structure $G$,  and $v_1v_2 \in G$,  $v_1v_3 \notin G$,  the output will be the state with graph structure $G'=G+v_1v_3$,  the weight of $v_1v_2$ in $G$ will be split by the new edge $v_1v_3$ in $G'$,  and the proportion is determined by $\theta$ in the controlled-$RBS(\theta)$ gate. So the controlled-$RBS(\theta)$ gate can be treated as ``edge generator”,  as long as the edge we want to add into the graph already has an adjacent edge in the original graph. On the other hand,  the inverse of the controlled-$RBS(\theta)$ gate is ``deleting" an edge (and transferring its amplitude to one of its adjacent edge) in $G$. Note that a controlled-$RBS(\theta)$ gate (and its inverse) has a constant size and depth.

In this subsection,  we prepare the $s\times t$-grid-structured state,  and denote the vertex in the $i^{th}$ row and the $j^{th}$ column as $v_{i, j}$,  and $q_{i, j}$ means the corresponding qubit. The pseudocode is shown in Algorithm \ref{Gridal}.

\begin{breakablealgorithm}
    \caption{The inverse circuit of preparing grid-structured states without ancillary qubits.}
    \begin{algorithmic}[1]
    \Require An arbitrary $s\times t$ grid $G=(V, E)$,  where $n=st=|V|$; $n$ working qubits.
    \Ensure The circuit that turns $\ket{\psi_G}$ into $\ket{0}^{\otimes n}$.
    \Procedure {GridCut}{$s$, $t$,  $s\times t$ grid $G$}
        \If{$t\ge 3$}
            \For{$k=1$ to $s$}
                \State Apply a inverse gate of the controlled-$RBS(\theta)$ gate on $q_{k, \lfloor t/2 \rfloor}, q_{k, \lfloor t/2 \rfloor+1}, q_{k, \lfloor t/2 \rfloor+2}$,  so that the edge  $v_{k, \lfloor t/2 \rfloor}v_{k, \lfloor t/2 \rfloor+1}$ is deleted.
            \EndFor
            \State We have cut $G$ into 2 small grids,  we name the left one $G_1$ and right one $G_2$.
            \State \Call{GridCut}{$s$, $\lfloor t/2 \rfloor$, $G_1$}
            \State \Call{GridCut}{$s$, $t-\lfloor t/2 \rfloor$, $G_2$}
        \ElsIf{$s\ge 3$}
            \State We do exactly the same work as line 3 to 4 by cutting $G$ into a top grid $G_1$ and a bottom grid $G_2$.
            \State \Call{GridCut}{$\lfloor s/2 \rfloor$, $t$, $G_1$}
            \State \Call{GridCut}{$s-\lfloor s/2 \rfloor$, $t$, $G_2$}
        \ElsIf{$t=s=2$}
            \State Delete $v_{(1, 1)}v_{(2, 1)}, v_{(2, 1)}v_{(2, 2)}, v_{(2, 2)}v_{(1, 2)}$ by applying 3 inverse gates of the controlled-$RBS(\theta)$ gates.
        \EndIf
        \State Return;
    \EndProcedure
    \State \nonumber
    \State \Call{GridCut}{$s$, $t$, $G$},  after which the final graph consists of independent edges.
    \State For every edge in the graph,  apply a CNOT gate controlled by one end and targeting on the other end.
    \State Apply the inverse circuit of the Unary Encoding circuit.
    \State Return the whole circuit.
    \end{algorithmic}
    \label{Gridal}
\end{breakablealgorithm}

\begin{lemma}
    Algorithm \ref{Gridal} correctly turns $\ket{\psi_G}$ into $\ket{0}^{\otimes n}$.
\end{lemma}

\noindent \textit{Proof.} Consider the operation of a single GridCut ($s$, $t$,  $s\times t$ grid $G$) function. If $t\ge 3$ or $s\ge 3$,  we simply remove all the edges in the middle. If $G$ is a $2\times 2$ grid,  we delete 3 edges so that there is only 1 edge left. As a result,  after line 19,  the final graph consists of independent edges. Assume the state after line 19 is of the form $\sum_{v_iv_j\in G_{\text{final}}}w_{ij}'\ket{e_i\oplus e_j}$. After line 20 the state becomes $\sum_{v_iv_j\in G_{\text{final}}}w_{ij}'\ket{e_j}$,  this is exactly the form of states after Unary Encoding. Hence we just apply the inverse of Unary Encoding (line 21) to get $\ket{0}^{\otimes n}$. For the case of complex coefficients, we need to change the controlled-$RBS(\theta)$ gate. For an arbitrary edge, we adjust its phase to the correct position in the last controlled-$RBS(\theta)$ gate that uses this edge as the ``generator''. $\hfill\square$

\begin{theorem}
    For an arbitrary grid $G$ with $n$ vertices,  $\ket{\psi_G}$ can be prepared by a circuit with $O(\log n)$-depth without ancillary qubits.
\end{theorem}

\noindent \textit{Proof.} In a single GridCut($s$, $t$,  $s\times t$ grid $G$) function,  if $t\ge 3$,  we simply remove all the edges in the middle. As different controlled-$RBS(\theta)$ gates work on different qubits,  line 3 and 4 only needs a depth $O(1)$. After we remove those edges,  we get 2 small grids,  $G_1,  G_2$. It is obvious that the number of vertices in each small grid is approximately $n/2$, strictly speaking, it is bounded above by $\frac{2}{3}n$. The number of vertices in each small grid is smaller than $\frac{2}{3}n$. As $G_1$ and $G_2$ are independent from each other,  line 11 and line 12 can work in parallel. The same analysis holds when $s\ge 3$. If we denote the depth of a GridCut function of an $n$-vertex grid as $f(n)$,  we get $f(n)\le O(1)+f(\frac{2}{3}n)$. So $f(n)=O(\log n)$.
As a result,  line 20 needs a depth of $O(\log n)$,  line 21 needs a depth of 1,  and line 19 needs a depth of $O(\log n)$,  so that the whole depth is $O(\log n)$.$\hfill\square$

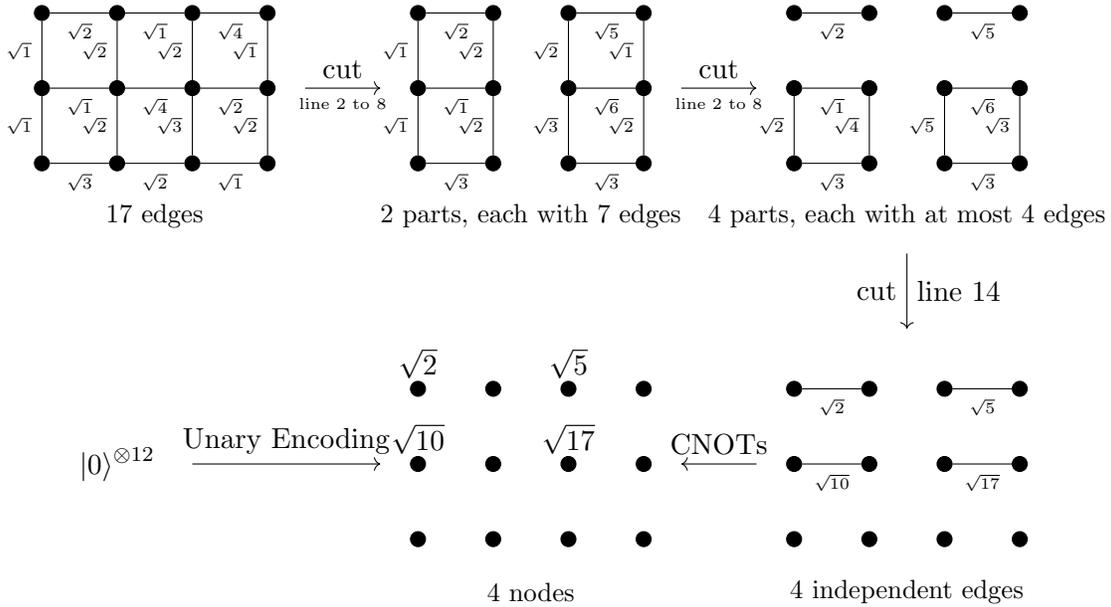
\begin{figure*}
    \centering
    \begin{tikzpicture}
        \node [circle,  draw,  fill=black,  inner sep=2pt](00) at (0, 0) {};
        \node [circle,  draw,  fill=black,  inner sep=2pt](10) at (1, 0) {};
        \node [circle,  draw,  fill=black,  inner sep=2pt](20) at (2, 0) {};
        \node [circle,  draw,  fill=black,  inner sep=2pt](30) at (3, 0) {};
        \node [circle,  draw,  fill=black,  inner sep=2pt](01) at (0, 1) {};
        \node [circle,  draw,  fill=black,  inner sep=2pt](11) at (1, 1) {};
        \node [circle,  draw,  fill=black,  inner sep=2pt](21) at (2, 1) {};
        \node [circle,  draw,  fill=black,  inner sep=2pt](31) at (3, 1) {};
        \node [circle,  draw,  fill=black,  inner sep=2pt](02) at (0, 2) {};
        \node [circle,  draw,  fill=black,  inner sep=2pt](12) at (1, 2) {};
        \node [circle,  draw,  fill=black,  inner sep=2pt](22) at (2, 2) {};
        \node [circle,  draw,  fill=black,  inner sep=2pt](32) at (3, 2) {};
        \node[]() at (1.5,-0.7){17 edges};

        \draw [->](3.5, 1)--node[above]{cut}node[below]{line 2 to 8}(4.5, 1);
        \draw (00)--node[below]{\tiny{$\sqrt{3}$}}(10);
        \draw (10)--node[below]{\tiny{$\sqrt{2}$}}(20);
        \draw (20)--node[below]{\tiny{$\sqrt{1}$}}(30);
        \draw (01)--node[below]{\tiny{$\sqrt{1}$}}(11);
        \draw (11)--node[below]{\tiny{$\sqrt{4}$}}(21);
        \draw (21)--node[below]{\tiny{$\sqrt{2}$}}(31);
        \draw (02)--node[below]{\tiny{$\sqrt{2}$}}(12);
        \draw (12)--node[below]{\tiny{$\sqrt{1}$}}(22);
        \draw (22)--node[below]{\tiny{$\sqrt{4}$}}(32);
        \draw (00)--node[left]{\tiny{$\sqrt{1}$}}(01);
        \draw (01)--node[left]{\tiny{$\sqrt{1}$}}(02);
        \draw (10)--node[left]{\tiny{$\sqrt{2}$}}(11);
        \draw (11)--node[left]{\tiny{$\sqrt{2}$}}(12);
        \draw (20)--node[left]{\tiny{$\sqrt{3}$}}(21);
        \draw (21)--node[left]{\tiny{$\sqrt{2}$}}(22);
        \draw (30)--node[left]{\tiny{$\sqrt{2}$}}(31);
        \draw (31)--node[left]{\tiny{$\sqrt{1}$}}(32);

        \begin{scope}[xshift=5cm, on grid]
        \node [circle,  draw,  fill=black,  inner sep=2pt](00) at (0, 0) {};
        \node [circle,  draw,  fill=black,  inner sep=2pt](10) at (1, 0) {};
        \node [circle,  draw,  fill=black,  inner sep=2pt](20) at (2, 0) {};
        \node [circle,  draw,  fill=black,  inner sep=2pt](30) at (3, 0) {};
        \node [circle,  draw,  fill=black,  inner sep=2pt](01) at (0, 1) {};
        \node [circle,  draw,  fill=black,  inner sep=2pt](11) at (1, 1) {};
        \node [circle,  draw,  fill=black,  inner sep=2pt](21) at (2, 1) {};
        \node [circle,  draw,  fill=black,  inner sep=2pt](31) at (3, 1) {};
        \node [circle,  draw,  fill=black,  inner sep=2pt](02) at (0, 2) {};
        \node [circle,  draw,  fill=black,  inner sep=2pt](12) at (1, 2) {};
        \node [circle,  draw,  fill=black,  inner sep=2pt](22) at (2, 2) {};
        \node [circle,  draw,  fill=black,  inner sep=2pt](32) at (3, 2) {};
        \node[]() at (1.5,-0.7){2 parts, each with 7 edges};

        \draw [->](3.5, 1)--node[above]{cut}node[below]{line 2 to 8}(4.5, 1);
        \draw (00)--node[below]{\tiny{$\sqrt{3}$}}(10);
        \draw (20)--node[below]{\tiny{$\sqrt{3}$}}(30);
        \draw (01)--node[below]{\tiny{$\sqrt{1}$}}(11);
        \draw (21)--node[below]{\tiny{$\sqrt{6}$}}(31);
        \draw (02)--node[below]{\tiny{$\sqrt{2}$}}(12);
        \draw (22)--node[below]{\tiny{$\sqrt{5}$}}(32);
        \draw (00)--node[left]{\tiny{$\sqrt{1}$}}(01);
        \draw (01)--node[left]{\tiny{$\sqrt{1}$}}(02);
        \draw (10)--node[left]{\tiny{$\sqrt{2}$}}(11);
        \draw (11)--node[left]{\tiny{$\sqrt{2}$}}(12);
        \draw (20)--node[left]{\tiny{$\sqrt{3}$}}(21);
        \draw (21)--node[left]{\tiny{$\sqrt{2}$}}(22);
        \draw (30)--node[left]{\tiny{$\sqrt{2}$}}(31);
        \draw (31)--node[left]{\tiny{$\sqrt{1}$}}(32);
        \end{scope}

        \begin{scope}[xshift=10cm, on grid]
        \node [circle,  draw,  fill=black,  inner sep=2pt](00) at (0, 0) {};
        \node [circle,  draw,  fill=black,  inner sep=2pt](10) at (1, 0) {};
        \node [circle,  draw,  fill=black,  inner sep=2pt](20) at (2, 0) {};
        \node [circle,  draw,  fill=black,  inner sep=2pt](30) at (3, 0) {};
        \node [circle,  draw,  fill=black,  inner sep=2pt](01) at (0, 1) {};
        \node [circle,  draw,  fill=black,  inner sep=2pt](11) at (1, 1) {};
        \node [circle,  draw,  fill=black,  inner sep=2pt](21) at (2, 1) {};
        \node [circle,  draw,  fill=black,  inner sep=2pt](31) at (3, 1) {};
        \node [circle,  draw,  fill=black,  inner sep=2pt](02) at (0, 2) {};
        \node [circle,  draw,  fill=black,  inner sep=2pt](12) at (1, 2) {};
        \node [circle,  draw,  fill=black,  inner sep=2pt](22) at (2, 2) {};
        \node [circle,  draw,  fill=black,  inner sep=2pt](32) at (3, 2) {};
        \node[]() at (1.5,-0.7){4 parts, each with at most 4 edges};
        
        \draw [->](1.5, -1.2)--node[left]{cut}node[right]{line 14}(1.5, -2.2);
        \draw (00)--node[below]{\tiny{$\sqrt{3}$}}(10);
        \draw (20)--node[below]{\tiny{$\sqrt{3}$}}(30);
        \draw (01)--node[below]{\tiny{$\sqrt{1}$}}(11);
        \draw (21)--node[below]{\tiny{$\sqrt{6}$}}(31);
        \draw (02)--node[below]{\tiny{$\sqrt{2}$}}(12);
        \draw (22)--node[below]{\tiny{$\sqrt{5}$}}(32);
        \draw (00)--node[left]{\tiny{$\sqrt{2}$}}(01);
        \draw (10)--node[left]{\tiny{$\sqrt{4}$}}(11);
        \draw (20)--node[left]{\tiny{$\sqrt{5}$}}(21);
        \draw (30)--node[left]{\tiny{$\sqrt{3}$}}(31);
        \end{scope}

        \begin{scope}[xshift=10cm, yshift=-5cm, on grid]
        \node [circle,  draw,  fill=black,  inner sep=2pt](00) at (0, 0) {};
        \node [circle,  draw,  fill=black,  inner sep=2pt](10) at (1, 0) {};
        \node [circle,  draw,  fill=black,  inner sep=2pt](20) at (2, 0) {};
        \node [circle,  draw,  fill=black,  inner sep=2pt](30) at (3, 0) {};
        \node [circle,  draw,  fill=black,  inner sep=2pt](01) at (0, 1) {};
        \node [circle,  draw,  fill=black,  inner sep=2pt](11) at (1, 1) {};
        \node [circle,  draw,  fill=black,  inner sep=2pt](21) at (2, 1) {};
        \node [circle,  draw,  fill=black,  inner sep=2pt](31) at (3, 1) {};
        \node [circle,  draw,  fill=black,  inner sep=2pt](02) at (0, 2) {};
        \node [circle,  draw,  fill=black,  inner sep=2pt](12) at (1, 2) {};
        \node [circle,  draw,  fill=black,  inner sep=2pt](22) at (2, 2) {};
        \node [circle,  draw,  fill=black,  inner sep=2pt](32) at (3, 2) {};
        \node[]() at (1.5,-0.7){4 independent edges};

        \draw (01)--node[below]{\tiny{$\sqrt{10}$}}(11);
        \draw (21)--node[below]{\tiny{$\sqrt{17}$}}(31);
        \draw (02)--node[below]{\tiny{$\sqrt{2}$}}(12);
        \draw (22)--node[below]{\tiny{$\sqrt{5}$}}(32);
        \end{scope}

        \begin{scope}[xshift=5cm, yshift=-5cm, on grid]
        \node [circle,  draw,  fill=black,  inner sep=2pt](00) at (0, 0) {};
        \node [circle,  draw,  fill=black,  inner sep=2pt](10) at (1, 0) {};
        \node [circle,  draw,  fill=black,  inner sep=2pt](20) at (2, 0) {};
        \node [circle,  draw,  fill=black,  inner sep=2pt](30) at (3, 0) {};
        \node [circle,  draw,  fill=black,  inner sep=2pt](01) at (0, 1) {};
        \node [circle,  draw,  fill=black,  inner sep=2pt](11) at (1, 1) {};
        \node [circle,  draw,  fill=black,  inner sep=2pt](21) at (2, 1) {};
        \node [circle,  draw,  fill=black,  inner sep=2pt](31) at (3, 1) {};
        \node [circle,  draw,  fill=black,  inner sep=2pt](02) at (0, 2) {};
        \node [circle,  draw,  fill=black,  inner sep=2pt](12) at (1, 2) {};
        \node [circle,  draw,  fill=black,  inner sep=2pt](22) at (2, 2) {};
        \node [circle,  draw,  fill=black,  inner sep=2pt](32) at (3, 2) {};
        \node[]() at (1.5,-0.7){4 nodes};

        \draw [<-](3.5, 1)--node[above]{CNOTs}(4.5, 1);
        \node[above] at (02) {$\sqrt{2}$};
        \node[above] at (01) {$\sqrt{10}$};
        \node[above] at (22) {$\sqrt{5}$};
        \node[above] at (21) {$\sqrt{17}$};
        
        \end{scope}

        \begin{scope}[yshift=-5cm, on grid]
        \node () at (1, 1) {$\ket{0}^{\otimes 12}$};
        \draw [->](2, 1)--node[above]{Unary Encoding}(4.5, 1);
        
        \end{scope}
    \end{tikzpicture} 
    \caption{An example of preparing a $3 \times 4$ grid-structured state.}
    \label{3*4}
\end{figure*}

Figure~\ref{3*4} is an example to show how Algorithm \ref{Gridal} works on a $3\times4$ grid. Starting from $\ket{\psi_G}$, we first apply GridCut(3, 4, $G$), so that we actually apply 3 inverse gates of controlled-$RBS(\theta)$ gates, and cut the entire grid into two small grids $G_1$ and $G_2$. Then we can apply GridCut(3, 2, $G_1$) and GridCut(3, 2, $G_2$) simultaneously. After 3 rounds of GridCut, we only have independent edges and nodes. For every edge, we apply a CNOT gate controlled by one end and targeting on the other end, so that the state we get is exactly the result of a Unary Encoding circuit.


\section{HWP states preparation in optimal depth}\label{sec:HWP}
While the preceding sections provided highly optimized solutions for the $k=2$ sector, many quantum algorithms and physical simulations—such as those involving multi-excitation manifolds in quantum chemistry—require the preparation of HWP states with arbitrary Hamming weights $k \ge 3$. As $k$ increases, the dimension of the target Hilbert space grows combinatorially as $D = \binom{n}{k}$, posing a significant challenge to maintaining low circuit depth. To push the boundaries of temporal efficiency, we generalize our framework in this section to address the $k \ge 3$ regime. When $k\ge 3$, we can still apply the circuit that we used in Theorem~\ref{thm:Graph Arbitrary}, but with a higher size.

\begin{lemma}\label{lem:weak HWP}
For any $n$-qubit HWP quantum state with $k$ excitations, it can be prepared by a circuit with $O\left(\log \binom{n}{k}\right)$ depth and $O\left(k \binom{n}{k}\right)$ size, utilizing $O\left(k \binom{n}{k}\right)$ ancillary qubits.
\end{lemma}

\noindent \textit{Proof.} This algorithm is quite similar to Algorithm \ref{Arbal}. We first apply the Unary Encoding on ${n \choose k}$ ancillary qubits. Each ancillary qubit $q'_x$ represents a computational basis in $\ket{\psi_{\text{H}}}$. This step needs ${n \choose k}$ ancillary qubits, and has $O\left(\log{{n \choose k}}\right)$-depth and $O\left({n \choose k}\right)$-size. We get the following transformation.
\begin{equation}
    \ket{0}^{\otimes n}\ket{0}^{\otimes {n \choose k}}\xrightarrow{\text{Unary Encoding}}\ket{0}^{\otimes n}\sum_{{\text{HW}(x)=k}}\alpha_x \ket{e_x}.
\end{equation}

Next, for each $q'_x$, we apply CNOT gates controlled by $q'_x$ and targeting on all the $q_i$s that the $i^{th}$ digit of $x$ is $1$. Because $\text{HW}(x)=k$, $k$ CNOT gates are applied here. As a result, there are $k{n \choose k}$ CNOT gates in total. This CNOT circuit is obviously a bipartite CNOT circuit. By Lemma~\ref{CNOTbipartite}, when using $k{n \choose k}$ ancillary qubits, this CNOT circuit can be implemented with $O\left(\log k{n \choose k}\right)=O\left(\log {n \choose k}\right)$-depth and $O\left(k{n \choose k}\right)$-size. We get the following transformation.

\begin{equation}
    \ket{0}^{\otimes n}\sum_{{\text{HW}(x)=k}}\alpha_x \ket{e_x}\xrightarrow{\text{CNOTs}}\sum_{{\text{HW}(x)=k}}\alpha_x \ket{x}\ket{e_x}.
\end{equation}

Finally, for each $q'_x$, we apply a $C^k{X}$ gate targeting on $q'_x$ and controlled by all the $q_i$s that the $i^{th}$ digit of $x$ is $1$. There are ${n \choose k}$ $C^k{X}$ gates in total. For each $q_i$, it is associated with ${n-1 \choose k-1}$ $C^k{X}$ gates. We make ${n-1 \choose k-1}$ copies of each $q_i$, so that all ${n \choose k}$ $C^k{X}$ gates can be applied simultaneously. Then we apply the inverse of the copying stage. In the copying (together with the inverse) stage, we make $n{n-1 \choose k-1}=k{n \choose k}$ copies in total, so we add $k{n \choose k}$ ancillary qubits and take $O\left(\log k{n \choose k}\right)=O\left(\log {n \choose k}\right)$-depth and $O\left(k{n \choose k}\right)$-size. Next in the $C^k{X}$ stage, by Lemma~\ref{lem:C^n(x)}, we add ${n \choose k}$ ancillary qubits and take $O(k)$-depth and $O\left(k{n \choose k}\right)$-size. We get the following transformation.

\begin{equation}
    \sum_{{\text{HW}(x)=k}}\alpha_x \ket{x}\ket{e_x}\xrightarrow{C^k (X) \text{s}}\left(\sum_{{\text{HW}(x)=k}}\alpha_x \ket{x}\right)\ket{0}^{\otimes {n \choose k}}.
\end{equation}

In aggregate, this circuit uses $O\left(k{n \choose k}\right)$ ancillary qubits, and has $O\left(\log{{n \choose k}}\right)$-depth and $O\left(k{n \choose k}\right)$-size. $\hfill\square$

Before detailing the optimal preparation algorithm, we first introduce a representation of the decimal number $x$ in Lemma \ref{lem:div}, and explain the necessary registers of ancillary qubits, which play key roles in the preparation algorithm design.

\begin{lemma}\label{lem:div}
    For all $x,i \in \mathbb{N}$ with $0 \leq x \leq 2^n-1$ and $1 \leq i \leq n$, there exists a unique ordered pair $(a, b)$ satisfying $a,b \in \mathbb{N}$ with $0\le a \le 2^{i}-1$ and $0 \le b \le 2^{n-i}-1$ so that $x=2^{n-i}a+b$. In quantum computation, this can be written as $\ket{x}=\ket{a}\ket{b}$.
\end{lemma}
\noindent \textit{Proof.} This follows straightforwardly from the division algorithm. $\hfill\square$

Next we consider the ancillary qubits, which have been separated into different registers. For simplicity, we assume that $k$ is an even number. 
Register $A$ contains ${n \choose k}$ ancillary qubits, and each qubit $q'_x$ represents a computational basis in $\ket{\psi_{\text{H}}}$. Register $B$ contains several small registers $B_{\frac{k}{2}}, B_{\frac{k}{2}+1}, \ldots B_{n-\frac{k}{2}}$, where $B_{i}$ contains ${i \choose \frac{k}{2}}+{n-i \choose \frac{k}{2}}$ ancillary qubits. In $B_{i}$, each qubit in the first ${i \choose \frac{k}{2}}$ qubits represents an $i$-bit string with Hamming Weight $\frac{k}{2}$, and each qubit in the last ${n-i \choose \frac{k}{2}}$ qubits represents a $(n-i)$-bit string with Hamming Weight $\frac{k}{2}$. Register $C$ contains several small registers $C_{\frac{k}{2}}, C_{\frac{k}{2}+1}, \ldots C_{n-\frac{k}{2}}$, where $C_{i}$ contains $n$ ancillary qubits. Registers $D,E,F$ each contains $n-k+1$ qubits respectively. Each qubit in register $D$ represents a position in an $n$-bit string, starting with the $\frac{k}{2}^{th}$ place and ending with the $(n-\frac{k}{2})^{th}$ place. Now we are ready to present the preparation algorithm for HWP states, as shown in Algorithm \ref{alg:HWP}.

\begin{breakablealgorithm}
    \caption{Preparation of a HWP state with $O\left({n \choose k}\right)$ ancillary qubits.}
    \begin{algorithmic}[1]
    \Require $n$; $k$; ${n \choose k}$ coefficients $\{\alpha_x\}$.
    \Ensure The circuit that prepares $\hwp$.
    \State Apply Unary Encoding on register $A$, where the coefficients are from $\{\alpha_x\}$.

    \For{every $x$ with $\text{HW}(x)=k$}
        \State Calculate the position of the $\frac{k}{2}^{th}$ $1$ in $x$(from left to right), denote as $i$. By Lemma~\ref{lem:div}, find $(a,b)$ so that $\ket{x}=\ket{a}\ket{b}$. By the choice of $i$, both $a$ and $b$ have a Hamming Weight of $\frac{k}{2}$.
        \State Apply 2 CNOT gates both controlled by $q_x'$ in register $A$, and targeting on the qubits representing $a$ and $b$ in register $B_i$.
        \State Apply 1 CNOT gate controlled by $q_x'$, and targeting on the qubit representing $i$ in register $D$.
    \EndFor

    \For{every edge $x$ with $\text{HW}(x)=k$}
        \State Apply a Toffoli gate controlled by the qubits representing $a$ and $b$ in register $B_i$, targeting on $q_x'$.
    \EndFor

    \For{every register $B_i$ in register $B$}
        \State Apply the circuit(except for the Unary Encoding step) introduced in Lemma~\ref{lem:weak HWP} to the first ${i \choose \frac{k}{2}}$ qubits and the last ${n-i \choose \frac{k}{2}}$ qubits respectively, and store the result in register $C_i$.
    \EndFor

    \For{every register $C_i$ in register $C$}
        \State Apply $n$ CNOT gates: for all $j\in [n]$, the $j^{th}$ CNOT gate is controlled by the $j^{th}$ qubit in $C_i$ and targeting on the working qubit $q_j$.
    \EndFor

    \For{every register $C_i$ in register $C$}
        \State Apply $n$ Toffoli gates: for all $j\in [n]$, the $j^{th}$ Toffoli gate is controlled by the working qubit $q_j$ and the qubit representing $i$ in register $D$, and targeting on the $j^{th}$ qubit in $C_i$.
    \EndFor

    \State Make $n-k+1$ copies of the working qubits and index them as: $\frac{k}{2},\frac{k}{2}+1,\ldots,n-\frac{k}{2}$.
    \For{$j$ from $\frac{k}{2}$ to $n-\frac{k}{2}$}
        \State Apply an HWC gate on the first $j$ qubit of the $j^{th}$ copy of the working qubits to check if the Hamming Weight equals $\frac{k}{2}$, and store the result in register $E$.
    \EndFor

    \State Apply an HWC gate on register $E$ to check if the Hamming Weight equals 1, and store the result in a single ancillary qubit $q^*$.

    \If {$q^*=1$}
        \State Apply $n-k+1$ CNOT gates: for all $j$ from $\frac{k}{2}$ to $n-\frac{k}{2}$, the $j^{th}$ CNOT gate is controlled by the $j^{th}$ qubit in $E$ and targeting on the $j^{th}$ qubit in $D$.
        \State All the operations in this situation should be controlled by $q^*$ when $q^*=1$.
    \Else
        \State Make $n-k+1$ copies of register $E$ and index them as: $\frac{k}{2},\frac{k}{2}+1,\ldots,n-\frac{k}{2}$.
        \For{$j$ from $\frac{k}{2}$ to $n-\frac{k}{2}$}
            \State Apply HWC on the first $j$ qubit of the $j^{th}$ copy of register $E$ to check if the Hamming Weight equals $1$, and store the result in register $F$.
        \EndFor
        \State Apply $n-k+1$ CNOT gates: for all $j$ from $\frac{k}{2}$ to $n-\frac{k}{2}$, the $j^{th}$ CNOT gate is controlled by the $j^{th}$ qubit in $F$ and targeting on the $j^{th}$ qubit in $D$.
        \State Apply the inverse circuit from line 28 to line 31.
        \State All the operations in this situation should be controlled by $q^*$ when $q^*=0$.
    \EndIf
    \State Apply the inverse circuit from line 19 to line 23.
    
    \State Return the whole circuit.
    \end{algorithmic}
    \label{alg:HWP}
\end{breakablealgorithm}

\begin{lemma}
    Algorithm \ref{alg:HWP} can correctly prepare the HWP state $\ket{\psi_{\text{H}}}$.
\end{lemma}

\noindent \textit{Proof.} We will explain how the initial state $\ket{0}^{\otimes n}$ can be transformed to the target HWP state. Line 1 is the Unary Encoding circuit. From line 2, we analyze the circuit by considering each possible input rather than the whole state for simplicity. The following is the transformation process of one possible input, which can be derived through straightforward computation. It is noteworthy that we may omit some unchanged ancillary registers and abbreviate the initial state to $\ket 0$ on them in the process.
\begin{scriptsize}
\begin{equation}
\begin{aligned}
     & \underbrace{\ket{0}^{\otimes n}}_{\text{working}}\underbrace{\ket{e_x}}_{A}\\
     \xrightarrow{\text{line 2 to 6}} & \underbrace{\ket{0}^{\otimes n}}_{\text{working}}\underbrace{\ket{e_x}}_{A}(\underbrace{\ket{0}}_{B_{\frac{k}{2}}}\ldots \underbrace{\ket{e_a}\ket{e_b}}_{B_i} \ldots \underbrace{\ket{0}}_{B_{n-\frac{k}{2}}})\underbrace{\ket{e_i}}_{D}\\
     \xrightarrow{\text{line 7 to 9}} & \underbrace{\ket{0}^{\otimes n}}_{\text{working}}\underbrace{\ket{0}}_{A}(\underbrace{\ket{0}}_{B_{\frac{k}{2}}}\ldots \underbrace{\ket{e_a}\ket{e_b}}_{B_i} \ldots \underbrace{\ket{0}}_{B_{n-\frac{k}{2}}})\underbrace{\ket{e_i}}_{D}\\
     \xrightarrow{\text{line 10 to 12}} & \underbrace{\ket{0}^{\otimes n}}_{\text{working}}(\underbrace{\ket{0}}_{B_{\frac{k}{2}}}\ldots \underbrace{\ket{0}\ket{0}}_{B_i} \ldots \underbrace{\ket{0}}_{B_{n-\frac{k}{2}}})(\underbrace{\ket{0}}_{C_{\frac{k}{2}}}\ldots \underbrace{\ket{x}}_{C_i} \ldots \underbrace{\ket{0}}_{C_{n-\frac{k}{2}}})\underbrace{\ket{e_i}}_{D}\\
     \xrightarrow{\text{line 13 to 15}} & \underbrace{\ket{x}}_{\text{working}}(\underbrace{\ket{0}}_{C_{\frac{k}{2}}}\ldots \underbrace{\ket{x}}_{C_i} \ldots \underbrace{\ket{0}}_{C_{n-\frac{k}{2}}})\underbrace{\ket{e_i}}_{D}\\
     \xrightarrow{\text{line 16 to 18}} & \underbrace{\ket{x}}_{\text{working}}(\underbrace{\ket{0}}_{C_{\frac{k}{2}}}\ldots \underbrace{\ket{0}}_{C_i} \ldots \underbrace{\ket{0}}_{C_{n-\frac{k}{2}}})\underbrace{\ket{e_i}}_{D}\\
     \xrightarrow{\text{discard }C} & \underbrace{\ket{x}}_{\text{working}}\underbrace{\ket{e_i}}_{D}\\
     \xrightarrow{\text{line 19}} & \underbrace{\ket{x}\ldots \ket{x}}_{(n-k+1) \text{times}}\underbrace{\ket{e_i}}_{D}.
\end{aligned}
\nonumber
\end{equation}
\end{scriptsize}
The next task is to restore the state in register $D$ to $\ket 0$. Based on the $(n-k+1)$ copies of $\ket x$, Line 20 to 22 checks the Hamming Weight of prefixes of $\ket{x}$, and register $E$ will be the form $\ket{0\ldots 0 1 \ldots 1 0 \ldots 0}$. The place of the first ``1" in register $E$ means the prefix ending with this place first has Hamming Weight $\frac{k}{2}$, so this place is exactly $i$ mentioned in line $3$. If there is only one ``1'' in register $E$, the value on $q^*$ will be $1$, and register $E$ equals register $D$. Register $D$ can be uncomputed by applying CNOT gates between the corresponding qubits in register $E$ and register $D$.


If there is more than one ``1'' in register $E$, the value on $q^*$ will be $0$. By checking the prefixes of register $E$, line 28 to 31 successfully removes the ``1"s after the first ``1" in register $E$ and put the new binary string into register $F$. The process continues by the following transformation:
\begin{small}
\begin{equation}
\begin{aligned}
     & \underbrace{\ket{x}\ldots \ket{x}}_{(n-k+1) \text{times}}\underbrace{\ket{e_i}}_{D}\\
     \xrightarrow{\text{line 20 to 22}} & \underbrace{\ket{x}\ldots \ket{x}}_{(n-k+1) \text{times}}\underbrace{\ket{e_i}}_{D}\underbrace{\ket{0\ldots 0 1 \ldots 1 0 \ldots 0}}_{E\text{ (more than one ``1'')}}\underbrace{\ket{0}}_{q^*}\\
     \xrightarrow{\text{line 23}} & \underbrace{\ket{x}\ldots \ket{x}}_{(n-k+1) \text{times}}\underbrace{\ket{e_i}}_{D}\underbrace{\ket{0\ldots 0 1 \ldots 1 0 \ldots 0}}_{E}\underbrace{\ket{0}}_{q^*}\\
    \xrightarrow{\text{line 28}} & \underbrace{\ket{x}\ldots \ket{x}}_{(n-k+1) \text{times}}\underbrace{\ket{e_i}}_{D}\underbrace{\ket{0\ldots 0 1 \ldots 1 0 \ldots 0}^{\otimes n-k+1}}_{\text{(n-k+1) copies of }E}\underbrace{\ket{0}}_{q^*}\\
    \xrightarrow{\text{line 29 to 31}} & \underbrace{\ket{x}\ldots \ket{x}}_{(n-k+1) \text{times}}\underbrace{\ket{e_i}}_{D}\underbrace{\ket{0\ldots 0 1 \ldots 1 0 \ldots 0}^{\otimes n-k+1}}_{\text{(n-k+1) copies of }E}\underbrace{\ket{e_i}}_{F}\underbrace{\ket{0}}_{q^*}\\
    \xrightarrow{\text{line 32}} & \underbrace{\ket{x}\ldots \ket{x}}_{(n-k+1) \text{times}}\underbrace{\ket{0}}_{D}\underbrace{\ket{0\ldots 0 1 \ldots 1 0 \ldots 0}^{\otimes n-k+1}}_{\text{(n-k+1) copies of }E}\underbrace{\ket{e_i}}_{F}\underbrace{\ket{0}}_{q^*}\\
     \xrightarrow{\text{line 33}} & \underbrace{\ket{x}\ldots \ket{x}}_{(n-k+1) \text{times}}\underbrace{\ket{0}}_{D}\underbrace{\ket{0\ldots 0 1 \ldots 1 0 \ldots 0}}_{E}\underbrace{\ket{0}}_{q^*}\\
     \xrightarrow{\text{line 36}} & \underbrace{\ket{x}}_{\text{working}}\underbrace{\ket{0}}_{D}\underbrace{\ket{0}}_{E}\underbrace{\ket{0}}_{q^*}.
\end{aligned}
\nonumber
\end{equation}
\end{small}
As a result, in both cases, the final output of the circuit will be $\ket{x}$, while all ancillary qubits are restored to $\ket{0}$. Thus, we finish the preparation of HWP states successfully. Also for the case of complex coefficients, we simply encode them in the Unary Encoding stage.
$\hfill\square$

\begin{theorem}
    For any $n$-qubit HWP quantum state with $k$ excitations, it can be prepared by a circuit with $O\left(\log{{n \choose k}}\right)$-depth using $O\left({n \choose k}\right)$ ancillary qubits.
\end{theorem}

\noindent \textit{Proof.} Line 1 is the Unary Encoding circuit, which has $O\left(\log {n \choose k}\right)$-depth and $O\left({n \choose k}\right)$-size.
From line 2 to 6, for every $x$ with $\text{HW}(x)=k$, 3 CNOT gates controlled by $q'_x$ are applied. There are $3{n \choose k}$ CNOT gates in this stage, and they form a bipartite CNOT circuit, by Lemma~\ref{CNOTbipartite}, when using $O\left({n \choose k}\right)$ ancillary qubits, this CNOT circuit can be implemented with $O\left(\log {n \choose k}\right)$-depth and $O\left({n \choose k}\right)$-size.

From line 7 to 9, for every $x$ with $\text{HW}(x)=k$, a Toffoli gate targeting on $q'_x$ is applied. There are ${n \choose k}$ Toffoli gates in this stage. By copying $2{n \choose k}$ qubits in register $B$, these Toffoli gates can be applied simultaneously. We further apply the inverse of the copying stage. Overall, when using $2{n \choose k}$ more ancillary qubits, this Toffoli circuit can be implemented with $O\left(\log {n \choose k}\right)$-depth and $O\left({n \choose k}\right)$-size.

From line 10 to 12, for every register $B_i$ in register $B$, by Lemma~\ref{lem:weak HWP}, the circuit depth for every $B_i$ is $O\left(\max{(\log {i \choose \frac{k}{2}},\log {n-i \choose \frac{k}{2}})}\right)$, which belongs to $O\left(\log {n \choose k}\right)$. Because the circuits for $B_i$s can be applied simultaneously, the overall depth of this stage is $O\left(\log {n \choose k}\right)$. Also by Lemma~\ref{lem:weak HWP}, the circuit size complexity for every $B_i$ is $O\left(\frac{k}{2}\left({i \choose \frac{k}{2}}+{n-i \choose \frac{k}{2}}\right)\right)$, hence the overall size complexity of this stage is $\sum_{i=\frac{k}{2}}^{n-\frac{k}{2}}O\left(\frac{k}{2}\left( {i \choose \frac{k}{2}}+ {n-i \choose \frac{k}{2}}\right)\right)$. By some combinatorial identities and the following estimation,

\begin{equation}\label{eq:bound}
\begin{aligned}
    & \sum_{i=\frac{k}{2}}^{n-\frac{k}{2}}\left(\frac{k}{2}\left({i \choose \frac{k}{2}}+{n-i \choose \frac{k}{2}}\right)\right) \\
     = & k{n-\frac{k}{2}+1 \choose \frac{k}{2}+1} \\
     \le & 4\left( \lceil \frac{k}{4} \rceil {n-\frac{k}{2}+1 \choose \frac{k}{2}+1}\right)\\
     \le & 4\left( \sum_{i=0}^{\lceil \frac{k}{4} \rceil-1} {n-\frac{k}{2}+1+i \choose \frac{k}{2}+1}\right)\\
     \le & 4 {n-\frac{k}{2}+\lceil \frac{k}{4} \rceil+1 \choose \frac{k}{2}+2}\\
     \le & 4 {n \choose k},
\end{aligned}
\end{equation}
the overall size complexity of this stage is $O\left({n \choose k}\right)$.

From line 13 to 15, there are $n(n-k)$ CNOT gates in total, and this is again a bipartite CNOT circuit; from line 16 to 18, there are $n(n-k)$ Toffoli gates in total. By the same method concerning line 2 to 6 and line 7 to 9, line 13 to 18 can be implemented with $O\left(\log {n \choose k}\right)$-depth and $O\left({n \choose k}\right)$-size.

Line 19 is a copying stage. We add $n(n-k)$ ancillary qubits, hence line 19 can be implemented with $O\left(\log n(n-k)\right)$-depth and $O\left(n(n-k)\right)$-size.

From line 20 to 22, we apply $n-k+1$ HWC gates simultaneously. By Lemma~\ref{lem:HWC}, when using $O\left(n^2(n-k+1)\right)$ ancillary qubits, this stage can be implemented with $O\left(\log n\right)$-depth and $O\left(n^2(n-k+1)\right)$-size.

Line 23 is trivial. Line 25, together with line 28 to 32 is also trivial by using the methods mentioned above. However, it should be noted that the operations in line 25 and line 28 to 32 are controlled by $q^*$. By Lemma~\ref{lem:C(U)}, they can still be implemented with $O\left(\log n\right)$-depth and $O\left(n^2(n-k+1)\right)$-size.

Line 36 is the inverse of some of the former operations and is therefore trivial.

We have analyzed the depth and size complexity of every stage in Algorithm~\ref{alg:HWP}, and all of them are upper bounded by $O\left(\log {n \choose k}\right)$-depth and $O\left({n \choose k}\right)$-size. The only thing that has not been mentioned is the number of ancillary qubits in register $B$, which is the only nontrivial case among all the registers. The number of ancillary qubits in register $B$ is $\sum_{i=\frac{k}{2}}^{n-\frac{k}{2}}\left(\log {i \choose \frac{k}{2}}+\log {n-i \choose \frac{k}{2}}\right)$. Actually, we have made a stronger estimation in Eq.~(\ref{eq:bound}), so the number of ancillary qubits in register $B$ is also upper bounded by $O\left({n \choose k}\right)$, which completes the proof. $\hfill\square$

The $O(\log \binom{n}{k})$-depth scheme represents a regime of maximal parallelism. However, a natural question arises regarding the spatial-temporal trade-off in HWP state preparation: can we minimize the spatial overhead if the depth complexity is not the primary constraint? We address this by exploring a resource-thrifty regime. We find that by sacrificing the logarithmic depth, the number of required ancillary qubits can be compressed to $O(\log k)$ without compromising the $O(\binom{n}{k})$ scaling of the total gate count, thereby maintaining optimality in terms of total quantum operation costs.

\section{HWP states preparation with fewer ancillary qubits}\label{sec:HWP fewer}

In this section, we introduce the algorithm for preparing a HWP state using $O\left(\log{k}\right)$ ancillary qubits. 
\begin{theorem}\label{thm:HWP con}
    For any $n$-qubit HWP quantum state with $k$ excitations, it can be prepared by a circuit with $O\left({n \choose k}\right)$-size using $O\left(\log{k}\right)$ ancillary qubits.
\end{theorem}

A natural framework for the quantum state preparation algorithm is applying several uniformly controlled gates (UCGs), first appeared in \cite{grover2002creating}. However, the size complexity of a general $n$-qubit diagonal unitary matrix is $O(2^n)$, which is prohibitively large for optimal state preparation. We need to improve the complexity making full use of the good properties of our aimed quantum states also based on this framework with UCGs.
A UCG on $n$ qubits is defined as 
\begin{equation}
    V_n=\text{diag}(U_0,U_1,\ldots,U_{2^{n-1}-1})\in\mathbb{C}^{2^n\times 2^n}.
\end{equation}
The operation of $V_n$ is 
\begin{equation}
    \ket{j}\ket{x_n}\xrightarrow{V_n}\ket{j}U_{j}\ket{x_n}, \forall~ 0\leq j \leq 2^{n-1}-1.
\end{equation}
It includes $2^{n-1}$ layers obviously, and each layer is a $(n-1)$-controlled gate, which conditions on every possible computational basis state $\ket k$ of the previous $n-1$ qubits and operates on the last qubit $\ket {x_n}$.

Following the approach in \cite{grover2002creating}, we employ a binary tree of height $n$ to represent a quantum state. The left branch means the value $0$ on a single node, and the right branch means the value $1$, hence the amplitudes of a quantum states can be stored in the leaf nodes of this binary tree, and every internal node stores the square root of sum of squares of its child nodes. Figure~\ref{fig:Comparison 42} shows the difference between binary trees of general $4$-qubit quantum states and $\ket{\psi^4_2}$. It is clear that when preparing $\ket{\psi^n_k}$, we only need to care about a small portion of branches, instead of all the branches, which means certain elements in the matrix can take arbitrary values, and thus we can simply disregard them. Still take $\ket{\psi^4_2}$ as an example, $V_4=\text{diag}(U_0,X,X,I,X,I,I,U_7)$, here $U_0$ and $U_7$ can be arbitrary unitary matrices, because after $V_1,V_2$ and $V_3$, computational basis states starting with $\ket{000\cdot}$ and $\ket{111\cdot}$ no longer exist (both of them can not make a computational base with Hamming Weight $2$ no matter the value on the last qubit). When decomposing $V_4$, we can find that among all the 16 elements in every diagonal unitary matrix, 4 of them can take arbitrary phases, which helps improve the size complexity.

\begin{figure*}
     \centering
\begin{tikzpicture}[
    level distance=12mm,  
    node/.style={circle, draw, minimum size=4pt, inner sep=0pt}, 
    edge/.style={thick}   
]

\node[node] (n00) at (0/2.5,0) {};

\node[node] (n10) at (-8/2.5,-1) {};
\node[node] (n11) at (8/2.5,-1) {};

\node[node] (n20) at (-12/2.5,-2) {};
\node[node] (n21) at (-4/2.5,-2) {};
\node[node] (n22) at (4/2.5,-2) {};
\node[node] (n23) at (12/2.5,-2) {};

\node[node] (n30) at (-14/2.5,-3) {};
\node[node] (n31) at (-10/2.5,-3) {};
\node[node] (n32) at (-6/2.5,-3) {};
\node[node] (n33) at (-2/2.5,-3) {};
\node[node] (n34) at (2/2.5,-3) {};
\node[node] (n35) at (6/2.5,-3) {};
\node[node] (n36) at (10/2.5,-3) {};
\node[node] (n37) at (14/2.5,-3) {};

\node[node] (n40) at (-15/2.5,-4) {}; 

\node[node] (n41) at (-13/2.5,-4) {};
\node[node] (n42) at (-11/2.5,-4) {};
\node[node, red] (n43) at (-9/2.5,-4) {};
\node[node] (n44) at (-7/2.5,-4) {};
\node[node, red] (n45) at (-5/2.5,-4) {};
\node[node, red] (n46) at (-3/2.5,-4) {};
\node[node] (n47) at (-1/2.5,-4) {};
\node[node] (n48) at (1/2.5,-4) {};
\node[node, red] (n49) at (3/2.5,-4) {};
\node[node, red] (n410) at (5/2.5,-4) {};
\node[node] (n411) at (7/2.5,-4) {};
\node[node, red] (n412) at (9/2.5,-4) {};
\node[node] (n413) at (11/2.5,-4) {};
\node[node] (n414) at (13/2.5,-4) {};
\node[node] (n415) at (15/2.5,-4) {};

\node[below, font=\tiny] at (n40.south) {$\ket{0000}$};  
\node[below, font=\tiny] at (n41.south) {$\ket{0001}$};  
\node[below, font=\tiny] at (n42.south) {$\ket{0010}$};  
\node[below, font=\tiny, red] at (n43.south) {$\ket{0011}$};  
\node[below, font=\tiny] at (n44.south) {$\ket{0100}$};  
\node[below, font=\tiny, red] at (n45.south) {$\ket{0101}$};  
\node[below, font=\tiny, red] at (n46.south) {$\ket{0110}$};  
\node[below, font=\tiny] at (n47.south) {$\ket{0111}$};  
\node[below, font=\tiny] at (n48.south) {$\ket{1000}$};  
\node[below, font=\tiny, red] at (n49.south) {$\ket{1001}$};  
\node[below, font=\tiny, red] at (n410.south) {$\ket{1010}$}; 
\node[below, font=\tiny] at (n411.south) {$\ket{1011}$}; 
\node[below, font=\tiny, red] at (n412.south) {$\ket{1100}$}; 
\node[below, font=\tiny] at (n413.south) {$\ket{1101}$}; 
\node[below, font=\tiny] at (n414.south) {$\ket{1110}$}; 
\node[below, font=\tiny] at (n415.south) {$\ket{1111}$}; 

\draw[edge, red] (n00) -- (n10);
\draw[edge, red] (n00) -- (n11);

\draw[edge, red] (n10) -- (n20);
\draw[edge, red] (n10) -- (n21);
\draw[edge, red] (n11) -- (n22);
\draw[edge, red] (n11) -- (n23);

\draw[edge] (n20) -- (n30);
\draw[edge, red] (n20) -- (n31);
\draw[edge, red] (n21) -- (n32);
\draw[edge, red] (n21) -- (n33);
\draw[edge, red] (n22) -- (n34);
\draw[edge, red] (n22) -- (n35);
\draw[edge, red] (n23) -- (n36);
\draw[edge] (n23) -- (n37);

\draw[edge] (n30) -- (n40);
\draw[edge] (n30) -- (n41);
\draw[edge] (n31) -- (n42);
\draw[edge, red] (n31) -- (n43);
\draw[edge] (n32) -- (n44);
\draw[edge, red] (n32) -- (n45);
\draw[edge, red] (n33) -- (n46);
\draw[edge] (n33) -- (n47);
\draw[edge] (n34) -- (n48);
\draw[edge, red] (n34) -- (n49);
\draw[edge, red] (n35) -- (n410);
\draw[edge] (n35) -- (n411);
\draw[edge, red] (n36) -- (n412);
\draw[edge] (n36) -- (n413);
\draw[edge] (n37) -- (n414);
\draw[edge] (n37) -- (n415);

\end{tikzpicture}
     \caption{General quantum states need full binary trees to represent, while $\ket{\psi^n_k}$ only need part of the binary tree (the red edges).}
     \label{fig:Comparison 42}
 \end{figure*}

Just like what we have done in the last example, we need to count the number of computational basis states that will exist in each $V_i$, which we will denote as $t_i$. In $V_i$, we need to consider the Hamming Weight on the first $i-1$ qubits. There are two concerns. In the first case, the Hamming Weight is too large (greater than $k$), which directly makes the Hamming Weight of the whole $n$-bit string greater than $k$ (no matter the values on the rest of the qubits). In the second case, the Hamming Weight is too small, which makes the Hamming Weight of the whole $n$-bit string less than $k$ (even if the values on the rest of the qubits are all $1$). Except for the above two cases, all the other computational basis will exist during the implementation of $V_i$. Due to the two concerns, all the UCGs can be categorized into three classes.

\begin{enumerate}
    \item When $1\le i\le k$, the Hamming Weight on the first $i-1$ qubits can take any number from $0$ to $i$, so $t_i=2\left({i-1 \choose 0}+{i-1 \choose 1}+\cdots+{i-1 \choose i}\right)=2^{i+1}$. Note that the factor $2$ means the value on the $i$th qubit could be either $0$ or $1$, and this case shows no difference compared to the original UCG construction;
    \item When $k+1\le i\le n-k+1$, the Hamming Weight on the first $i-1$ qubits can take any number from $0$ to $k$, so $t_i=2\left({i-1 \choose 0}+{i-1 \choose 1}+\cdots+{i-1 \choose k}\right)$;
    \item When $n-k+2\le i\le n$, the Hamming Weight on the first $i-1$ qubits can take any number from $i-(n-k+1)$ to $k$, so $t_i=2\left({i-1 \choose i-(n-k+1)}+{i-1 \choose i-(n-k)}+\cdots+{i-1 \choose k}\right)$.
\end{enumerate}

In order to match the size lower bound of this problem, we add $\text{HWC}^i_k$ before $V_i$ when we meet case $2$ and case $3$, this $\text{HWC}^i_k$ is controlled by the first $i-1$ qubits and targets on a new ancillary qubit. $V_i$ will then be $0$-controlled by this ancillary qubits. At last, we add the inverse of $\text{HWC}^i_k$. This step simply prevent computational basis on the first $i-1$ qubits with Hamming Weight $k$ from $V_i$. This is still a correct circuit because originally, computational basis on the first $i-1$ qubits with Hamming Weight $k$ will be applied an $I$ gate, which means there's no difference whether they will exist in $V_i$.

Hence we need to recount the number of computational basis states that exist in $V_i$, by deleting the last summand in case $2$ and case $3$, we get 

\begin{enumerate}
    \item When $1\le i\le k$, the Hamming Weight on the first $i-1$ qubits can take any number from $0$ to $i$, so $t_i=2\left({i-1 \choose 0}+{i-1 \choose 1}+\cdots+{i-1 \choose i-1}\right)=2^{i+1}$. Note that the factor $2$ means the value on the $i$th qubit could be either $0$ or $1$, and this case shows no difference compared to the original UCG construction;
    \item When $k+1\le i\le n-k+1$, the Hamming Weight on the first $i-1$ qubits can take any number from $0$ to $k$, so $t_i=2\left({i-1 \choose 0}+{i-1 \choose 1}+\cdots+{i-1 \choose k-1}\right)$;
    \item When $n-k+2\le i\le n$, the Hamming Weight on the first $i-1$ qubits can take any number from $i-(n-k+1)$ to $k$, so $t_i=2\left({i-1 \choose i-(n-k+1)}+{i-1 \choose i-(n-k)}+\cdots+{i-1 \choose k-1}\right)$.
\end{enumerate}

We will synthesize each $V_i$ so that the size complexity scales linearly with $t_i$. Therefore, the size complexity of our problem is the sum of all $t_i$s, which can be written as the sum of many binomial coefficients. Through some combinatorial identities,

\begin{equation}
\begin{aligned}
    \sum_{i=1}^n t_i & =2\sum_{j=0}^{k-1}\sum_{i=j}^{j+n-k} {i \choose j}\\
    & =2\sum_{j=0}^{k-1} {j+n-k+1 \choose j+1}\\
    & =2\left({n+1 \choose k}-1\right)\\
    & <4{n \choose k}.
\end{aligned}
\end{equation}

As a result, the size complexity of this circuit is $O\left({n \choose k}\right)$.

\subsection{Synthesis of UCGs}

The core of our synthesis strategy is by sequentially implementing diagonal unitary matrices and single-qubit gates. Crucially, Lemma \ref{lem:Z-Ydecom}, in conjunction with subsequent analysis, jointly establishes the functional equivalence.

\begin{lemma}\label{lem:Z-Ydecom}
    ($Z-Y$ decomposition for a single qubit \cite{nielsen2010quantum}) Any $2\times 2$ unitary matrix $U$ can be decomposed as products of $R_y$ and $R_z$ gates. More specifically, there exist real numbers $\alpha, \beta, \gamma$ and $\delta$, such that
    \begin{equation}
        U=e^{i\alpha}R_z(\beta)R_y(\gamma)R_z(\delta).
    \end{equation}
\end{lemma}
Simple arithmetic demonstrates that $R_y(\gamma)=SHR_z(\gamma)HS^{\dagger},$ hence yielding
\begin{equation}\label{eq:single_qubit_gate}
    U=e^{i\alpha}R_z(\beta)SHR_z(\gamma)HS^{\dagger}R_z(\delta).
\end{equation}

And any UCG can be decomposed into
\begin{widetext}
\begin{multline} \label{eq:UCG}
    {V_n}=\underbrace{
    {\text{diag}(e^{i\alpha_1},\ldots,e^{i\alpha_{2^{n-1}}})\otimes \mathbb{I}_1}
    }_{A}  \cdot \underbrace{
   {\text{diag}(R_z(\beta_1),\ldots,R_z(\beta_{2^{n-1}}))
    }}_{B}\\\cdot
    \underbrace{  {\mathbb{I}_{n-1}\otimes (SH)}}_{C}
     \cdot\underbrace{
    {\text{diag}(R_z(\gamma_1),\ldots,R_z(\gamma_{2^{n-1}}))}
    }_{D}\cdot
     \underbrace{ { \mathbb{I}_{n-1}\otimes (HS^\dagger)}}_{E}
     \cdot\underbrace{
    {\text{diag}(R_z(\delta_1),\ldots,R_z(\delta_{2^{n-1}}))}
    }_{F}.
\end{multline}
\end{widetext}
The operator $C$ and $E$ can be applied by adding single-qubit gates on the last qubit($SH$ and $HS^\dagger$ respectively); $A,B,D$ and $F$ are all diagonal unitary matrices, which means they only have non-zero elements on their main diagonals and those elements are all complex numbers with norm $1$.

As mentioned before, there are computational basis that will not exist in $V_i$, which means for $A,B,D$ and $F$, only $t_i$ elements on each main diagonal are required to take specific values, and other elements can take arbitrary values. Next we show the algorithm that implements a diagonal unitary matrix with this restriction. Obviously, a diagonal unitary matrix only changes the phases of computational bases, and we can add an ancillary qubit and apply $R_z$ gates on it. For every computational basis that exists in $V_i$, denoted as a binary string $\ket{x}=\ket{x_1x_2\ldots x_i}$, for example, we apply a CNOT gate controlled by $q_j$, targeting on the ancilla, whenever $x_j=1$. Then we apply an $R_z$ gate on the ancillary qubit, the parameter $\theta_x$ will be determined later. At last we apply the inverse of those CNOT gates. We call the whole three small steps an oracle for $\ket{x}$. After we apply oracles for all computational bases that exist in $V_i$, we have successfully finished the synthesis of the diagonal unitary matrix. There are two questions that remain unanswered about this algorithm, and we address them in turn.

\begin{itemize}
    \item Why is this algorithm correct? How to determine the parameters of these $R_z$ gates?
    \item What's the size complexity of this algorithm?
\end{itemize}

We consider the first question. When some computational basis $\ket{y}$ enters the oracle for $\ket{x}$, it will get a phase shift due to $R_z(\theta_x)$. Specifically, the phase shift is $(-1)^{\langle x,y \rangle}\theta_x$, here we simply assume that $\langle \cdot, \cdot \rangle$ is over $\mathbb{F}_2$. As a result, as an input of the diagonal unitary matrix, $\ket{y}$ will get a phase shift of $\sum_{x \text{ that exists}}(-1)^{\langle x,y \rangle}\theta_x$, and it has to meet the requirement. This forms a linear equation system of parameters of $R_z$ gates. We will consider this problem in the last subsection.

Then we consider the second question. Applying these oracles directly is highly inefficient. A simple example can help understand the claim. When we subsequently apply oracles for $\ket{1110}$ and $\ket{1111}$, the third step in the first oracle needs $3$ CNOTs, and $4$ CNOTS in the first step of the second oracle. Among the 7 CNOTs, 6 of them can be deleted directly, which shows huge waste in original design. Instead, we can simply use $1$ CNOT to finish the two small steps, simply because the Hamming Distance between $1110$ and $1111$ is $1$. Inspired by this example, we can design a binary string sequence that consecutive strings differ by Hamming Distance less than a constant number.

\begin{theorem}
    There exists a sequence $S_{n,k}$ of all $n$-bit binary strings with Hamming Weight $k$ such that any two consecutive strings differ in exactly two bits, and the sequence starts with $\underbrace{1 \ldots 1}_{k \text{ times}} \underbrace{0 \ldots 0}_{(n-k) \text{ times}}$ and ends with string $\underbrace{0 \ldots 0}_{(n-k) \text{ times}} \underbrace{1 \ldots 1}_{k \text{ times}}$.
\end{theorem}

\noindent \textit{Proof.} We prove by induction on $n$ and $k$ together. When $k=1$, the claim is trivial, by simply shifting the only $1$ from left to right. For larger $k$, we apply $S_{n-1,k-1}$ on the rightmost $n-1$ bits. We have put all $n$-bit binary strings that start with $1$ into the sequence, and the currently last string is $1\underbrace{0 \ldots 0}_{(n-k-1) \text{ times}} \underbrace{1 \ldots 1}_{(k-1) \text{ times}}$. We then shift the leftmost $1$ to the second place, and then apply the inverse of $S_{n-2,k-1}$ on the rightmost $n-2$ bits. After that, we have put all $n$-bit binary strings that start with $01$ into the sequence. Again we shift the leftmost $1$ to the right position (note that at this situation, the third bit is already $1$, we put the leftmost $1$ to the $k+2^{th}$ place). The whole sequence $S_{n,k}$ is generated by repeatedly applying the above process. $\hfill\square$

Back to the second question, now we can design the binary string sequence. Assume that the Hamming Weight limit on the first $i-1$ qubits is $p\le \text{HW} \le q$. We start from $\underbrace{1 \ldots 1}_{p \text{ times}} \underbrace{0 \ldots 0}_{(i-p-1) \text{ times}}0$ and apply $S_{i-1,p}$, note that the last qubit could either be $0$ or $1$, before each step, we need to flip the last bit of the binary string and then flip it back(this is a constant number of operations). After this, the sequence temporarily ends with $ \underbrace{0 \ldots 0}_{(i-p-1) \text{ times}}\underbrace{1 \ldots 1}_{p \text{ times}}0$ and we have finished all the binary strings which have the Hamming Weight $p$ on the first $i-1$ qubits. Then we flip the $i-p-1$th digit, making the Hamming Weight increase by $1$, and apply the inverse of $S_{i-1,p+1}$. The whole sequence can be generated by repeating the above process.

Now we can calculate the size complexity. As consecutive strings differ by Hamming Distance no more than 3. The size complexity is at most $(3+1) t_i=4t_i$, here $+1$ means the $R_z$ gate in each oracle.

As a result, when synthesizing $V_i$, we need to implement 4 diagonal unitary matrices and 2 single-qubit gates, hence the overall complexity is $O(t_i)$.

\subsection{The existence of parameters in $R_z$ gates}

\begin{definition}\label{def:A}
    Let $A(n,k_1,k_2)$ be the matrix with rows and columns indexed by $x, y \in \{0, 1\}^n$ such that $k_1 \le \text{HW}(x), \text{HW}(y) \le k_2$, defined by:
    \begin{equation}
        A_{x,y} = (-1)^{|x \cap y|},
    \end{equation}
    where $|x \cap y| = \sum_{i=1}^n x_i y_i$.
\end{definition}

\begin{theorem}
If either $k_1=0$ or $k_2-k_1$ is odd, $A(n,k_1,k_2)$ is invertible.
\end{theorem}
\noindent\textit{Proof.} The proof is given in the Appendix.

Back to the first question, as the computational bases in $V_i$ have a Hamming Weight limit on the first $i-1$ qubits(and could either be $0$ or $1$ on the $i$th qubit), the coefficient matrix of parameters of $R_z$ gates is the tensor product of $A(i-1,k_1,k_2)$ and $H$ for some $k_1,k_2$. In case 1 and 2, $k_1=0$, hence $A(i-1,k_1,k_2)$ is invertible. In case 3, $k_2=k-1$ but $k_1=i-(n-k+1)>0$. As a result, $k_2-k_1$ may not be odd in this case. To fix this problem, we introduce a smaller Hamming Weight layer when $k_2-k_1$ is even, which means $k^{*}_1=k_1-1$ becomes the new second parameter in $A(i-1,k_1,k_2)$, this action guarantees that $A(i-1,k^{*}_1,k_2)$ is again invertible, but increases the size complexity. $t^{*}_i\le t_i+ 2{i-1 \choose i-(n-k+1)-1}$. Fortunately,
\begin{equation}
\begin{aligned}
    \sum_{i=1}^n t^{*}_i & \le \sum_{i=1}^n t^{}_i+2\sum_{n-k+2}^n {{i-1 \choose i-(n-k+1)-1}}\\
    & <4{n \choose k}+2{n \choose k-2}\\
    & <6{n \choose k}.
\end{aligned}
\end{equation}
By the above adjustment, there must be a set of solutions to the equation system of parameters of $R_z$ gates in every $V_i$, and we have insured the correctness of this algorithm.

\section{Lower Bound}\label{sec:lower}
The algorithms described in the preceding sections demonstrate remarkable versatility in balancing spatial and temporal resources. However, the performance of any quantum state preparation scheme is ultimately constrained by the amount of information required to specify the target state and the time necessary for entanglement to propagate across the system. To rigorously evaluate the standing of our work within the broader context of quantum complexity, it is essential to determine the theoretical floor for these resource costs. The following theorem provides these fundamental benchmarks, confirming that our constructions reach the absolute frontiers of circuit efficiency.

\begin{theorem}
For any $n$ and $k$, the preparation of a generic HWP state $\ket{\psi_{\text{H}}}$ using a universal gate set requires a circuit of size $\Omega\left(\binom{n}{k}\right)$ and depth $\Omega\left(\log \binom{n}{k}\right)$, regardless of the number of available ancillary qubits.
\end{theorem}
\noindent \textit{Proof.} We first consider depth complexity. For a circuit $C$ with $n$ working qubits and $l$ ancillary qubits, 
we suppose it has a depth of $D$ to prepare the objective $\ket\psi$.

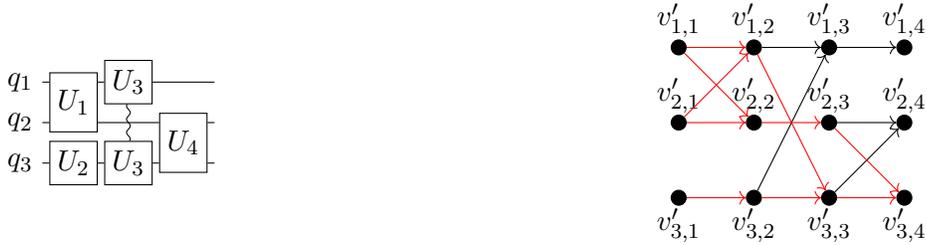
\begin{figure}[h]
    \centering
    \begin{subfigure}[c]{0.45\textwidth}
    \centering
    \begin{tikzpicture}
 \begin{yquant}
qubit {$q_1$} a;
qubit {$q_2$} a[+1];
qubit {$q_3$} a[+1];
 box {$U_{1}$} (-a[1]);
 box {$U_2$} (a[2]);
 box {$U_{3}$} (a[0, 2]);
 box {$U_{4}$} (a[1]-);
 \end{yquant}
 \end{tikzpicture}
    \end{subfigure}
\hfill
    \begin{subfigure}[c]{0.45\textwidth}
    \centering
\begin{tikzpicture}
        \node [circle,  draw,  fill=black,  inner sep=2pt](00) at (0, 0) {};
        \node [circle,  draw,  fill=black,  inner sep=2pt](10) at (1, 0) {};
        \node [circle,  draw,  fill=black,  inner sep=2pt](20) at (2, 0) {};
        \node [circle,  draw,  fill=black,  inner sep=2pt](30) at (3, 0) {};
        \node [circle,  draw,  fill=black,  inner sep=2pt](01) at (0, 1) {};
        \node [circle,  draw,  fill=black,  inner sep=2pt](11) at (1, 1) {};
        \node [circle,  draw,  fill=black,  inner sep=2pt](21) at (2, 1) {};
        \node [circle,  draw,  fill=black,  inner sep=2pt](31) at (3, 1) {};
        \node [circle,  draw,  fill=black,  inner sep=2pt](02) at (0, 2) {};
        \node [circle,  draw,  fill=black,  inner sep=2pt](12) at (1, 2) {};
        \node [circle,  draw,  fill=black,  inner sep=2pt](22) at (2, 2) {};
        \node [circle,  draw,  fill=black,  inner sep=2pt](32) at (3, 2) {};

        \node[below] at (00) {$v'_{3, 1}$};
        \node[above] at (01) {$v'_{2, 1}$};
        \node[above] at (02) {$v'_{1, 1}$};
        \node[below] at (10) {$v'_{3, 2}$};
        \node[above] at (11) {$v'_{2, 2}$};
        \node[above] at (12) {$v'_{1, 2}$};
        \node[below] at (20) {$v'_{3, 3}$};
        \node[above] at (21) {$v'_{2, 3}$};
        \node[above] at (22) {$v'_{1, 3}$};
        \node[below] at (30) {$v'_{3, 4}$};
        \node[above] at (31) {$v'_{2, 4}$};
        \node[above] at (32) {$v'_{1, 4}$};

        \draw [->, red](00)--(10);
        \draw [->, red](01)--(11);
        \draw [->, red](02)--(12);
        \draw [->, red](01)--(12);
        \draw [->, red](02)--(11);
        \draw [->, red](10)--(20);
        \draw [->](10)--(22);
        \draw [->, red](12)--(20);
        \draw [->](12)--(22);
        \draw [->, red](20)--(30);
        \draw [->](20)--(31);
        \draw [->, red](21)--(30);
        \draw [->](21)--(31);
        \draw [->, red](11)--(21);
        \draw [->](22)--(32);
\end{tikzpicture}
    \end{subfigure}

    \caption{A simple quantum circuit and its corresponding directed graph. The red lines shows the light cone of $q_3$.}
    \label{lightcone}
\end{figure}

Next, we define a new {\em directed} graph $G'=\{V', E'\}$. $V'$,  the vertex set of $G'$,  can be divided into $D+1$ layers, denoted as $v'_{i,j} $ ($i\in [n+l], j\in [D+1]$), representing the current state. Each layer contains $n+l$ vertices,  corresponding to all working and ancillary qubits. As a result,  $|V'|=(D+1)(n+l)$. Edges only appear between two vertices in adjacent layers. The starting (ending) vertex of an edge is always the vertex in the previous (latter) layer. $v'_{i,k}$ in the $k^{th}$ layer and $v'_{j,k+1}$ in the $(k+1)^{th}$ layer are adjacent if and only if $q_i$ and $q_j$ join the same gate in the $k^{th}$ layer of the circuit. Note that if some $q_i$ does not change in the $k^{th}$ layer,  it means there is an identity gate on $q_i$ in the $k^{th}$ layer,  so $v'_{i,k}$ in the $k^{th}$ layer and $v'_{i,k+1}$ in the $(k+1)^{th}$ layer must be adjacent.  
Obviously,  a single-qubit gate in the circuit will generate $1$ edge in the circuit,  while a double-qubit gate will generate $4$ edges. Figure~\ref{lightcone} shows a simple circuit and its corresponding directed graph. 

Here we define the light cone of $G'$. The light cone of $G'$ is a subset of $V'$. A vertex $v'$ belongs to the light cone if and only if there exists a directed path that starts from $v'$ and ends in one of the working vertices in the last layer, that is some $v'_i$ in the $(D+1)^{th}$ layer, $i\in\{1, 2, \dots, n\}$. It is not hard to see that only gates within the light cone contribute to the generation of $\ket{\psi}$. Thus,  we can simply ignore the gates outside the cone! (The effect of the removal is that the final state may change into $\ket{\psi}\ket{\phi}$ for some $\ket{\phi}$ different from $\ket{0}^{\otimes l}$.)

Actually, any vertex $v'_i$ in the $(k+1)^{th}$ layer can be the end vertex of at most two directed edges. This leads to the fact that the light cone will contain at most $O(n\cdot 2^D)$ vertices. As each gate can be specified by $O(1)$ parameters,  the circuit (after deleting all the gates outside the light cone) can generate at most $O(n\cdot 2^D)$ parameters. 

Since an HWP state is characterized by ${n \choose k}$ coefficients, the set of all such states with fixed $n$ and $k$ constitutes a ${n \choose k}$-dimensional vector space.
If $D\le \frac{1}{2}\log{{n \choose k}}$, the number of parameters generated by the $D$-depth circuit $C$ will be $o\left({n \choose k}\right)$.
When the light cone contains $o\left({n \choose k}\right)$ parameters, the set of HWP states implementable by any circuit sharing $C$'s structure will be contained within a measure-zero set of the vector space. And there are only finitely distinct circuit structures at depth $D$ , thus the collection of all HWP states that can be prepared within depth $D$ equals the union of finitely many measure-zero sets, which is still measure-zero.
Consequently, almost all of the HWP state with parameters $n$ and $k$ can not be prepared by a circuit of depth $\frac{1}{2}\log{n \choose k}$.


Next, we consider size complexity, and this is easier. On the one hand, since each single- or double-qubit gate can be specified by $O(1)$ parameters, the preparation circuit with size $S$ has at most $O(S)$ parameters. On the other hand, an HWP state has ${n \choose k}$ coefficients. Similar to the depth analysis, to generate the HWP state successfully, the number of parameters in the circuit must be no less than that of coefficients in the HWP state. Thus, almost all the HWP states with parameters $n$ and $k$ need to be prepared in $\Omega\left({n \choose k}\right)$-size. $\hfill\square$

\section{Numerical Results}\label{sec:numerical}
\begin{figure}
    \centering
    \includegraphics[width=\linewidth]{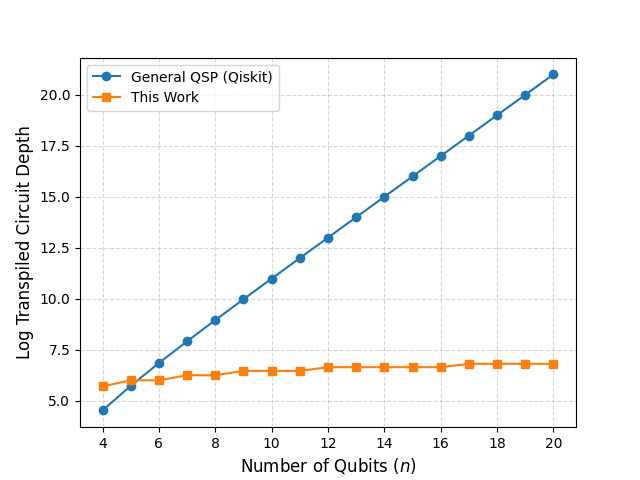}
    \caption{Numerical benchmarking of HWP state synthesis for 1D $XX$ model.}
    \label{fig:exp}
\end{figure}

To benchmark the practical performance of our HWP synthesis protocol, we investigate the digital preparation of Bethe states. We consider the one-dimensional $XX$ model, which serves as the non-interacting limit ($\Delta=0$) of the $XXZ$ Heisenberg spin chain. The Hamiltonian under open boundary conditions (OBC) is defined as:
\begin{equation}
H = J \sum_{i=1}^{n-1} \left( \sigma_i^x \sigma_{i+1}^x + \sigma_i^y \sigma_{i+1}^y \right),
\end{equation}
where $J$ is the exchange coupling and $\sigma^{x,y}$ are Pauli operators. Due to the $U(1)$ symmetry of the system, the total excitation number is conserved, and the eigenstates (Bethe states) naturally reside within the HWP sectors. We focus on the $k=2$ sector, where the target Bethe state $\ket{\psi_{\text{B}}}=\sum\alpha_{i,j}\ket{e_i \oplus e_j}$encodes critical information about many-body correlations. By mapping the system to free fermions, the coefficients $\alpha_{i, j}$ for each basis state $\ket{e_i \oplus e_j}$ (where $1 \le i < j \le n$) are analytically determined by the Slater determinant of the two lowest-energy sine modes, i.e.,
\begin{small}
\begin{equation}
\mathcal{N} \left[ \sin\left(\frac{\pi i}{n+1}\right)\sin\left(\frac{2\pi j}{n+1}\right) - \sin\left(\frac{\pi j}{n+1}\right)\sin\left(\frac{2\pi i}{n+1}\right) \right],
\end{equation}
\end{small}
\noindent ensures the unit normalization of the state. In our numerical experiment, we compare our proposed synthesis algorithm with the standard StatePreparation protocol provided by Qiskit \cite{iten2016quantum,aleksandrowicz2019qiskit}. Both circuits are transpiled into one-qubit gates and CNOT gates to provide a rigorous comparison of hardware resources. As illustrated in the scatter plot (Fig.~\ref{fig:exp}), the general-purpose Qiskit baseline exhibits a steep scaling in circuit depth as the system size $n$ increases. In contrast, our HWP-specific algorithm consistently maintains a logarithmic depth of $\Theta(\log n)$ while preserving an optimal circuit size of $\Theta(m)$. The significant reduction in depth directly translates to higher fidelity in near-term quantum simulations of many-body magnetism, where decoherence remains a primary bottleneck.

\section{Conclusion and Outlook}\label{sec:conclusion}

In summary, we have developed an asymptotically optimal framework for Hamming-Weight-Preserving (HWP) state preparation. For general excitation number $k$, our construction achieves $\Theta(\log \binom{n}{k})$ circuit depth and $\Theta(\binom{n}{k})$ circuit size, matching the corresponding information-theoretic lower bounds. We further provide an ancilla-efficient scheme requiring only $O(\log k)$ ancillary qubits. For the $k=2$ sector, we show that states whose support graphs are trees or grids admit ancilla-free preparation with $\Theta(\log n)$ depth and $\Theta(m)$ size. Together, these results establish efficient preparation protocols across both general and structured HWP states.

Several directions remain open for future investigation. First, it would be interesting to characterize possible tradeoffs among ancillary qubit count, circuit size, and circuit depth, with the goal of further reducing the ancilla requirement. Second, the circuit optimization techniques developed in this work, including CNOT circuit processing and qubit copying assignment, may also be useful in other quantum circuit synthesis and state preparation tasks. Finally, our ancilla-efficient construction relies on the non-singularity of a class of matrices associated with Krawtchouk polynomials. A complete characterization of the spectra and algebraic structure of these matrices remains open and may reveal further connections between quantum state preparation and Hamming-space structures.

\appendix
\section{Invertibility of $A(n,k_1,k_2)$}

Let
\begin{equation}
H=
\begin{pmatrix}
1&1\\
1&-1
\end{pmatrix}.
\end{equation}
The rows and columns of $H^{\otimes n}$ are indexed by subsets
$S,T\subseteq[n]$, equivalently by binary strings in $\{0,1\}^n$.
Under this identification,
\begin{equation}
(H^{\otimes n})_{S,T}
=
(-1)^{|S\cap T|}.
\end{equation}

For integers $0\le k_1\le k_2\le n$, define $A(n,k_1,k_2)$ to be the
principal submatrix of $H^{\otimes n}$ indexed by all subsets whose
cardinalities lie in $[k_1,k_2]$. Thus
\begin{equation}
A(n,k_1,k_2)_{S,T}
=
(-1)^{|S\cap T|},
\qquad
k_1\le |S|,|T|\le k_2.
\end{equation}

Our goal is to prove the following theorem.

\begin{theorem}
Let $0\le k_1\le k_2\le n$. If either $k_1=0$ or $k_2-k_1$ is odd, $A(n,k_1,k_2)$ is invertible.
\end{theorem}

\subsection{Symmetric-group decomposition}

For $0\le k\le n$, let
\begin{equation}
X_k=\{S\subseteq[n]:|S|=k\},
\qquad
V_k=\C^{X_k},
\end{equation}
and define
\begin{equation}
V_{[k_1,k_2]}
=
\bigoplus_{k=k_1}^{k_2}V_k.
\end{equation}

The symmetric group $S_n$ acts on $V_k$ by
\begin{equation}
\sigma e_S=e_{\sigma S}.
\end{equation}
Since $|\sigma S\cap\sigma T|=|S\cap T|$, we have
\begin{equation}
A(n,k_1,k_2)\sigma
=
\sigma A(n,k_1,k_2)
\end{equation}
for every $\sigma\in S_n$. Hence
\begin{equation}
A(n,k_1,k_2)
\in
\End_{S_n}\left(V_{[k_1,k_2]}\right).
\end{equation}

For $1\le j\le n$, define the lowering operator
$D_j:V_j\to V_{j-1}$ by
\begin{equation}
D_je_J
=
\sum_{\substack{R\subset J\\|R|=j-1}}e_R.
\end{equation}
For $0\le j\le \lfloor n/2\rfloor$, define the harmonic subspace
\begin{equation}
\mathcal H_j=\ker D_j.
\end{equation}

The permutation module $V_k$ has the multiplicity-free decomposition \cite{prasad2015representation,etingof2011introduction}
\begin{equation}
V_k
\cong
\bigoplus_{j=0}^{\min(k,n-k)}
S^{(n-j,j)}.
\end{equation}
Moreover,
\begin{equation}
\mathcal H_j\cong S^{(n-j,j)},
\end{equation}
and
\begin{equation}
\dim\mathcal H_j
=
\binom nj-\binom n{j-1},
\end{equation}
where $\binom n{-1}=0$\cite{dunkl1976krawtchouk}.

For $j\le k\le n-j$, define
$E_{j,k}:\mathcal H_j\to V_k$ by
\begin{equation}
(E_{j,k}h)(S)
=
\sum_{\substack{J\subseteq S\\|J|=j}}h(J).
\end{equation}
Then $E_{j,k}(\mathcal H_j)$ is the copy of $S^{(n-j,j)}$ contained in
$V_k$. This is the natural extension of a harmonic function from rank $j$
to the $k$-th fiber of the Boolean lattice, in the framework of
discrete harmonics on regular lattices; see \cite{delsarte1978hahn}.

For each $j$, define
\begin{equation}
I_j
=
\left\{
k\in\mathbb Z:
\max(k_1,j)\le k\le \min(k_2,n-j)
\right\}.
\end{equation}
Only those $j$ for which $I_j\neq\varnothing$ occur in the decomposition.
Thus
\begin{equation}
V_{[k_1,k_2]}
\cong
\bigoplus_{\substack{
0\le j\le \lfloor n/2\rfloor\\
I_j\neq\varnothing
}}
\mathcal H_j\otimes\C^{I_j}.
\end{equation}

Since $A(n,k_1,k_2)$ commutes with $S_n$, Schur's lemma implies that on
the $j$-th isotypic component it has the form
\begin{equation}
I_{\mathcal H_j}\otimes B_j,
\end{equation}
where $B_j$ is a matrix indexed by $I_j$. Therefore
\begin{equation}
A(n,k_1,k_2)\text{ is invertible}
\iff
B_j\text{ is invertible for every }j.
\end{equation}

\subsection{Explicit form of the reduced blocks}
AI assistance was used in developing the proofs presented in the following three subsections.

For $k,\ell\in[k_1,k_2]$, let $M_{k,\ell}:V_\ell\to V_k$ be the
corresponding block of $A(n,k_1,k_2)$:
\begin{equation}
(M_{k,\ell}f)(S)
=
\sum_{|T|=\ell}
(-1)^{|S\cap T|}f(T).
\end{equation}

Fix $j$ and suppose $j\le k,\ell\le n-j$. For $h\in\mathcal H_j$,
\begin{equation}
\begin{aligned}
(M_{k,\ell}E_{j,\ell}h)(S)
&=
\sum_{|T|=\ell}
(-1)^{|S\cap T|}
\sum_{\substack{J\subseteq T\\|J|=j}}h(J)
\\
&=
\sum_{|J|=j}h(J)
\sum_{\substack{T\supseteq J\\|T|=\ell}}
(-1)^{|S\cap T|}.
\end{aligned}
\end{equation}

For fixed $S,J$, put $a=|S\cap J|$. Since $|S|=k$, the inner sum depends
only on $a$. Denote it by $q(a)$. We have
\begin{equation}
q(a)
=
(-1)^a
\sum_t
(-1)^t
\binom{k-a}{t}
\binom{n-k-j+a}{\ell-j-t}.
\end{equation}
Equivalently,
\begin{equation}
q(a)
=
[z^{\ell-j}]
(-1)^a
(1-z)^{k-a}
(1+z)^{n-k-j+a},
\end{equation}
and hence
\begin{equation}
q(a)
=
[z^{\ell-j}]
(1-z)^k(1+z)^{n-k-j}
\left(
-\frac{1+z}{1-z}
\right)^a.
\end{equation}

Write the Newton expansion
\begin{equation}
q(a)
=
\sum_{r=0}^{j}c_r\binom ar,
\qquad
c_r=\Delta^rq(0).
\end{equation}

Since $h\in\mathcal H_j$, for every $R\subseteq[n]$ with $|R|=r<j$,
\begin{equation}
\sum_{\substack{|J|=j\\J\supseteq R}}h(J)=0.
\end{equation}
Indeed, the coefficient of $e_R$ in
$D_{r+1}D_{r+2}\cdots D_jh$ is a nonzero constant multiple of this sum,
while
\begin{equation}
D_{r+1}D_{r+2}\cdots D_jh=0.
\end{equation}

Consequently, for $r<j$,
\begin{equation}
\begin{aligned}
\sum_{|J|=j}
h(J)\binom{|S\cap J|}{r}
&=
\sum_{\substack{R\subseteq S\\|R|=r}}
\sum_{\substack{|J|=j\\J\supseteq R}}h(J)
\\
&=0.
\end{aligned}
\end{equation}
For $r=j$,
\begin{equation}
\sum_{|J|=j}
h(J)\binom{|S\cap J|}{j}
=
\sum_{\substack{J\subseteq S\\|J|=j}}h(J)
=
(E_{j,k}h)(S).
\end{equation}
It follows that
\begin{equation}
M_{k,\ell}E_{j,\ell}h
=
c_jE_{j,k}h.
\end{equation}

Now
\begin{equation}
\begin{aligned}
c_j
&=
\Delta^jq(0)
\\
&=
[z^{\ell-j}]
(1-z)^k(1+z)^{n-k-j}
\left(
-\frac{1+z}{1-z}-1
\right)^j
\\
&=
(-2)^j
[z^{\ell-j}]
(1-z)^{k-j}
(1+z)^{n-k-j}.
\end{aligned}
\end{equation}

Define the binary Krawtchouk polynomials \cite{krawtchouk1929generalisation, delsarte1973algebraic, feinsilver2005krawtchouk} by
\begin{equation}
K_s^{(N)}(x)
=
[z^s](1-z)^x(1+z)^{N-x}.
\end{equation}
Then
\begin{equation}
c_j
=
(-2)^j
K_{\ell-j}^{(n-2j)}(k-j).
\end{equation}
Therefore
\begin{equation}
(B_j)_{k,\ell}
=
(-2)^j
K_{\ell-j}^{(n-2j)}(k-j).
\end{equation}

For integers $0\le L\le U\le N$, define
\begin{equation}
C_N[L,U]
=
\left(
K_s^{(N)}(r)
\right)_{r,s=L}^{U}.
\end{equation}

For every $j$ with $I_j\neq\varnothing$, put
\begin{equation}
N_j=n-2j,\
L_j=\max(k_1,j)-j,
\ U_j=\min(k_2,n-j)-j.
\end{equation}
Then $0\le L_j\le U_j\le N_j$ and
\begin{equation}
B_j=(-2)^jC_{N_j}[L_j,U_j].
\end{equation}

\subsection{Krawtchouk orthogonality}

\begin{lemma}
For $0\le s,t\le N$,
\begin{equation}
\sum_{x=0}^{N}
\binom Nx
K_s^{(N)}(x)K_t^{(N)}(x)
=
2^N\binom Ns\delta_{st}.
\end{equation}
\end{lemma}

\begin{proof}
Using the generating function
\begin{equation}
K_s^{(N)}(x)
=
[z^s](1-z)^x(1+z)^{N-x},
\end{equation}
we compute
\begin{equation}
\begin{aligned}
&\sum_{x=0}^{N}
\binom Nx
(1-z)^x(1+z)^{N-x}
(1-w)^x(1+w)^{N-x}
\\
&=
\sum_{x=0}^{N}
\binom Nx
\bigl((1-z)(1-w)\bigr)^x
\bigl((1+z)(1+w)\bigr)^{N-x}
\\
&=
\left(
(1-z)(1-w)+(1+z)(1+w)
\right)^N
\\
&=
2^N(1+zw)^N.
\end{aligned}
\end{equation}
Taking the coefficient of $z^sw^t$ gives
\begin{equation}
\sum_{x=0}^{N}
\binom Nx
K_s^{(N)}(x)K_t^{(N)}(x)
=
2^N\binom Ns\delta_{st}.
\end{equation}
\end{proof}

The polynomial $K_s^{(N)}(x)$ has degree exactly $s$. Therefore
$K_0^{(N)},K_1^{(N)},\ldots,K_d^{(N)}$ form a basis of the space of
polynomials of degree at most $d$.

Define
\begin{equation}
\langle f,g\rangle
=
\sum_{x=0}^{N}
\binom Nx f(x)\overline{g(x)}.
\end{equation}
The orthogonality relation implies that if
$P\in\Span\{K_L^{(N)},\ldots,K_U^{(N)}\}$, then
\begin{equation}
\langle P,R\rangle=0
\end{equation}
for every polynomial $R$ satisfying $\deg R<L$.

\subsection{Three invertibility lemmas for \(C_N[L,U]\)}

\begin{lemma}
For every $0\le U\le N$, the matrix $C_N[0,U]$ is invertible.
\end{lemma}

\begin{proof}
Suppose that $C_N[0,U]$ is singular. Then there exist coefficients
$a_0,\ldots,a_U$, not all zero, such that
\begin{equation}
\sum_{s=0}^{U}a_sK_s^{(N)}(r)=0,
\qquad
r=0,1,\ldots,U.
\end{equation}
Define
\begin{equation}
P(x)
=
\sum_{s=0}^{U}a_sK_s^{(N)}(x).
\end{equation}
Since $\deg K_s^{(N)}=s$, we have $\deg P\le U$. On the other hand,
\begin{equation}
P(0)=P(1)=\cdots=P(U)=0.
\end{equation}
Thus $P$ has $U+1$ distinct zeros while having degree at most $U$.
Hence $P\equiv0$. Since
$K_0^{(N)},\ldots,K_U^{(N)}$ are linearly independent, all $a_s$ must
vanish, a contradiction. Therefore $C_N[0,U]$ is invertible.
\end{proof}

\begin{lemma}
For every $1\le L\le N$, the matrix $C_N[L,N]$ is invertible.
\end{lemma}

\begin{proof}
Suppose that $C_N[L,N]$ is singular. Then there exist coefficients
$a_L,\ldots,a_N$, not all zero, such that
\begin{equation}
\sum_{s=L}^{N}a_sK_s^{(N)}(r)=0,
\qquad
r=L,L+1,\ldots,N.
\end{equation}
Define
\begin{equation}
P(x)
=
\sum_{s=L}^{N}a_sK_s^{(N)}(x).
\end{equation}
Then
\begin{equation}
P(L)=P(L+1)=\cdots=P(N)=0.
\end{equation}
Hence
\begin{equation}
P(x)=Q(x)R(x),
\qquad
Q(x)=\prod_{r=L}^{N}(x-r).
\end{equation}

Since $\deg P\le N$ and $\deg Q=N-L+1$, we have
\begin{equation}
\deg R\le L-1<L.
\end{equation}
By Krawtchouk orthogonality,
\begin{equation}
\langle P,R\rangle=0.
\end{equation}
Therefore
\begin{equation}
0
=
\sum_{x=0}^{N}
\binom Nx Q(x)|R(x)|^2.
\end{equation}

For $x=L,\ldots,N$, we have $Q(x)=0$. For every integer
$x=0,\ldots,L-1$, all $N-L+1$ factors in $Q(x)$ are negative. Hence
\begin{equation}
\operatorname{sgn}Q(x)
=
(-1)^{N-L+1},
\end{equation}
independently of $x$, and $Q(x)\neq0$.

Thus all nonzero terms in
\begin{equation}
\sum_{x=0}^{N}
\binom Nx Q(x)|R(x)|^2
\end{equation}
have the same sign. Since the sum is zero,
\begin{equation}
R(0)=R(1)=\cdots=R(L-1)=0.
\end{equation}
The polynomial $R$ has at least $L$ distinct zeros but degree at most
$L-1$. Therefore $R\equiv0$. Hence $P\equiv0$, contradicting the
linear independence of $K_L^{(N)},\ldots,K_N^{(N)}$. Therefore
$C_N[L,N]$ is invertible.
\end{proof}

\begin{lemma}
Let $1\le L\le U\le N$. If $U-L$ is odd, then $C_N[L,U]$ is
invertible.
\end{lemma}

\begin{proof}
Suppose that $C_N[L,U]$ is singular. Then there exist coefficients
$a_L,\ldots,a_U$, not all zero, such that
\begin{equation}
\sum_{s=L}^{U}a_sK_s^{(N)}(r)=0,
\qquad
r=L,L+1,\ldots,U.
\end{equation}
Define
\begin{equation}
P(x)
=
\sum_{s=L}^{U}a_sK_s^{(N)}(x).
\end{equation}
Then
\begin{equation}
P(L)=P(L+1)=\cdots=P(U)=0.
\end{equation}
Hence
\begin{equation}
P(x)=Q(x)R(x),
\qquad
Q(x)=\prod_{r=L}^{U}(x-r).
\end{equation}

Since $\deg P\le U$ and $\deg Q=U-L+1$, we obtain
\begin{equation}
\deg R
\le
U-(U-L+1)
=
L-1<L.
\end{equation}
By Krawtchouk orthogonality,
\begin{equation}
\langle P,R\rangle=0.
\end{equation}
Therefore
\begin{equation}
0
=
\sum_{x=0}^{N}
\binom Nx Q(x)|R(x)|^2.
\end{equation}

Since $U-L$ is odd, $U-L+1$ is even. For every integer $x<L$, all
factors of $Q(x)$ are negative and their number is even, so $Q(x)>0$.
For every integer $x>U$, all factors are positive, so again $Q(x)>0$.
Moreover,
\begin{equation}
Q(x)=0,
\qquad
L\le x\le U.
\end{equation}
Therefore $Q(x)\ge0$ for every $x\in\{0,1,\ldots,N\}$, with strict
positivity outside $[L,U]$.

It follows that every term in
\begin{equation}
\sum_{x=0}^{N}
\binom Nx Q(x)|R(x)|^2
\end{equation}
is nonnegative. Since the sum equals zero,
\begin{equation}
R(x)=0
\end{equation}
for every
$x\in\{0,\ldots,L-1\}\cup\{U+1,\ldots,N\}$. In particular,
\begin{equation}
R(0)=R(1)=\cdots=R(L-1)=0.
\end{equation}
Thus $R$ has at least $L$ distinct zeros, while $\deg R\le L-1$.
Hence $R\equiv0$. Therefore $P\equiv0$, contradicting the linear
independence of $K_L^{(N)},\ldots,K_U^{(N)}$. Thus $C_N[L,U]$ is
invertible.
\end{proof}

\subsection{Proof of the main theorem}

\begin{proof}[Proof of the Theorem]
For every $j$ with $I_j\neq\varnothing$, we have
\begin{equation}
B_j=(-2)^jC_{N_j}[L_j,U_j],
\end{equation}
where
\begin{equation}
N_j=n-2j,
\
L_j=\max(k_1,j)-j,
\
U_j=\min(k_2,n-j)-j.
\end{equation}
It is therefore sufficient to prove that every
$C_{N_j}[L_j,U_j]$ is invertible.

First suppose $k_1=0$. Then
\begin{equation}
L_j
=
\max(0,j)-j
=
0
\end{equation}
for every $j$. Hence every reduced block is of the form
$C_{N_j}[0,U_j]$, which is invertible by the first invertibility lemma.
Therefore $A(n,0,k_2)$ is invertible.

Now suppose that $k_2-k_1$ is odd. Fix $j$ such that
$I_j\neq\varnothing$.

If $j\ge k_1$, then $L_j=0$. Hence
$C_{N_j}[L_j,U_j]=C_{N_j}[0,U_j]$ is invertible.

It remains to consider $j<k_1$. In this case
\begin{equation}
L_j=k_1-j>0.
\end{equation}

If $k_2<n-j$, then $U_j=k_2-j$, and therefore
\begin{equation}
U_j-L_j
=
(k_2-j)-(k_1-j)
=
k_2-k_1,
\end{equation}
which is odd. Hence $C_{N_j}[L_j,U_j]$ is invertible by the third
invertibility lemma.

If instead $k_2\ge n-j$, then
\begin{equation}
U_j
=
n-j-j
=
n-2j
=
N_j.
\end{equation}
Thus the reduced block is $C_{N_j}[L_j,N_j]$, which is invertible by
the second invertibility lemma.

Therefore every reduced block $B_j$ is invertible. By the
$S_n$-isotypic decomposition, $A(n,k_1,k_2)$ is invertible.
\end{proof}

\begin{acknowledgments}
AI tools were used to assist in developing parts of the proofs presented in the Appendix.
This work was supported in part by the National Natural Science Foundation of China Grants No. 62325210, Quantum Science and Technology-National Science and Technology Major Project under Grant No. 2024ZD0300500, and Beijing Natural Science Foundation No. 4252013.
\end{acknowledgments}

\bibliographystyle{apsrev4-2}
\bibliography{ref1}

\end{document}